\documentclass[english,11pt,a4paper]{article}
\pdfoutput=1
\usepackage{jheppub}
\usepackage{axodraw2}
\usepackage{amsmath}
\usepackage{amsfonts}
\usepackage{amssymb}
\usepackage{mathtools}
\usepackage{pifont} 
\usepackage{color}
\usepackage{bm}		
\usepackage{graphicx}
\usepackage{feynmp}
\usepackage{slashed}	
\usepackage{xspace}	
\usepackage{subcaption}
\usepackage{epigraph}
\usepackage{braket}
\usepackage[table,xcdraw]{xcolor}

\providecommand{\beq}{\begin{equation}}
\providecommand{\eeq}{\end{equation}}

\providecommand{\nl}{\nonumber \\}

\newcommand{\eqnref}[1]{Eq.~(\ref{eqn:#1})}

\newcommand{\secref}[1]{Sec.~\ref{sec:#1}}

\newcommand{\appref}[1]{Appendix~\ref{app:#1}}
\newcommand{\figref}[1]{Fig.~\ref{fig:#1}}

\DeclareMathOperator\Order{\mathcal{O}}
\providecommand{\mean}[1]{\left\langle #1 \right\rangle}

\newcommand\GeV{\, \mathrm{GeV}} 
\newcommand\eff{\mathrm{eff}} 
\renewcommand\max{\mathrm{max}} 
\newcommand\BO{\mathrm{BO}} 
\newcommand\ann{\mathrm{ann}} 
\newcommand\rc{R_c} 
\newcommand\xc{x_c} 
\newcommand\kL{\kappa_{\Lambda_D}} 
\newcommand\XXb{$X$--$\bar{X}$\xspace} 

\newcommand\bfP{{\boldsymbol{P}}}
\newcommand\bfp{{\boldsymbol{p}}}
\newcommand\bfK{{\boldsymbol{K}}}
\newcommand\bfk{{\boldsymbol{k}}}
\newcommand\bfr{{\boldsymbol{r}}}
\newcommand\bfR{{\boldsymbol{R}}}

\newcommand\bfx{{\boldsymbol{x}}}

\newcommand\bfz{{\boldsymbol{z}}}

\newcommand\bfF{{\boldsymbol{F}}}
\newcommand\bfe{{\boldsymbol{\epsilon}}}

\newcommand\bfPphi{\bfP_\varphi}
\newcommand\bfcalM{{\boldsymbol{\mathcal M}}}

\def\e{\epsilon}
\def\f{\phi}

\def\j{\psi}

\def\O{\Omega}

\newcommand{\ThreeJ}[6]{\begin{pmatrix}#1 & #2 & #3 \\ #4 & #5 & #6\end{pmatrix}}
\newcommand{\ThreeJzero}[3]{\ThreeJ{#1}{#2}{#3}000}
\newcommand\w[1]{_{\mathrm{#1}}}
\newcommand\fracp[3]{\left(\frac{#1}{#2}\right)^{#3}}
\newcommand{\intdMom}{\int\!\!\dMom}
\newcommand{\dMom}[2]{\frac{d^{#1}#2}{(2\pi)^{#1}}}

\title{Dark quarkonium formation in the early universe}

\author[a]{M.~Geller,}
\author[b,c]{S.~Iwamoto,}
\author[b,d,e]{G.~Lee,}
\author[b]{Y.~Shadmi,}
\author[d]{and O.~Telem}

\affiliation[a]{Maryland Center for Fundamental Physics, Department of Physics, University of Maryland, College Park, MD 20742, USA}
\affiliation[b]{Physics Department, Technion---Israel Institute of Technology, Haifa 32000, Israel}
\affiliation[c]{Dipartimento di Fisica e Astronomia, Universit\`a di Padova, Via Marzolo~8, I-35131 Padua, Italy}
\affiliation[d]{Department of Physics, LEPP, Cornell University, Ithaca, NY 14853, USA}
\affiliation[e]{Department of Physics, Korea University, Seoul 136--713, Republic of Korea}

\emailAdd{mic.geller@gmail.com}
\emailAdd{sho.iwamoto@pd.infn.it}
\emailAdd{gabr.lee@cornell.edu}
\emailAdd{yshadmi@physics.technion.ac.il}
\emailAdd{t10ofrit@gmail.com}

\abstract{%
  The relic abundance of heavy stable particles charged under a confining gauge
  group can be depleted by a second stage of annihilations near the deconfinement temperature. 
  This proceeds via the formation of quarkonia-like states, in which the heavy pair subsequently annihilates.
  The size of the quarkonium formation cross section was the subject of some debate.
  We estimate this cross section in a simple toy model.
  The dominant process can be viewed as a rearrangement of the heavy and
  light quarks, leading to a geometric cross section of hadronic size.
  In contrast, processes in which only the heavy constituents are involved
  lead to mass-suppressed cross sections.
  These results apply to any scenario with bound states of sizes much larger
  than their inverse mass,
  such as U(1) models
  with charged particles of different masses,
and can be used to construct ultra-heavy dark-matter models with masses above the na\"ive unitarity bound.
They are also relevant for the cosmology of any stable colored relic.
}

\begin{document}
\maketitle



\section{Introduction} \label{sec:intro}

\epigraph{... and in the darkness bind them.}{\textit{J.\,R.\,R.\,Tolkien}}

\noindent%
Stable ``colored'' particles, charged under QCD or a hidden confining gauge group, have been proposed as dark matter (DM) candidates~\cite{Jacoby:2007nw, Berger:2008ti, Cheung:2008ke, Chen:2009ab,Davoudiasl:2010am,Feng:2011ik,Blinov:2012hq,Bai:2013xga,Cline:2013zca,Boddy:2014yra,Cline:2014kaa, Boddy:2014qxa,Forestell:2016qhc,Harigaya:2016nlg,Cline:2016nab,Asadi:2016ybp,Liew:2016hqo,Cirelli:2016rnw,Cline:2017tka,Forestell:2017wov,DeLuca:2018mzn,Berlin:2018tvf},
and are predicted in various extensions of the Standard Model~\cite{Baer:1998pg, Kang:2006yd, Arvanitaki:2005fa}.
Even in the simplest models, the cosmological history of colored relics
is intriguing, and their present-day abundances have been the subject of some debate.
The relic abundance of a heavy colored particle $X$ is sensitive to the two inherent scales in the problem:
its mass $m_X$, and the confinement scale $\Lambda_D$.
If $m_X\gg\Lambda_D$, the freeze-out of $X$ proceeds
via standard perturbative annihilations 
at temperatures $T \sim m_X/30$.
However, at temperatures $T\sim\Lambda_D$, long after the perturbative \XXb annihilations
have shut off, the $X$ relic abundance may be further reduced by interactions of hadronized $X$s,
whose size is set by $1/\Lambda_D$.

The annihilation process at $T \lesssim \Lambda_D$ was described in a semiclassical approximation in Ref.~\cite{Kang:2006yd}.
At $T \sim \Lambda_D$, most of the $X$s are in color-singlet heavy-light hadrons, which we label by $H_X$.
An $X$-hadron $H_X$ and an $\bar{X}$-hadron $\bar{H}_X$ experience a residual strong interaction whose effective range is $\sim 1/\Lambda_D$,
much larger than the Compton wavelength $\sim 1/m_X$, but much smaller than the mean distance between $X$-hadrons at $T \sim \Lambda_D$. 
\XXb annihilation then proceeds via the formation of \XXb ``quarkonia'',
which subsequently de-excite to the ground state.
In the ground state, the \XXb distance is of order the Compton wavelength,
and the pair annihilates into light mesons, (dark) photons, or glueballs.
In the following, we use parentheses to denote quarkonium-like states, and
refer to the $(X\bar{X})$ states as quarkonia.

The cross section for quarkonia formation 
was argued in Ref.~\cite{Kang:2006yd} to be purely geometric.
This certainly holds for the scattering cross section of two $H_X$ hadrons;
however, it is less clear that it holds for the quarkonium production
cross section, which requires a significant modification of the trajectories of the heavy particles.
One semiclassical argument for a mass-suppressed cross section was
described in Ref.~\cite{Arvanitaki:2005fa}.
To form a bound state, the $X$ and $\bar{X}$ must lose energy and angular momentum. 
Classically, one can estimate the cross section by modeling the energy loss as Larmor radiation. 
This is proportional to the acceleration-squared, which scales as $\Lambda_D^4/m_X^2$.
At $T\sim \Lambda_D$, the $X$s are very slow, with speed $v\sim \sqrt{\Lambda_D/m_X}$.
Thus it takes a long time for the hadrons to cross a distance $1/\Lambda_D$;
however, the total amount of radiation still scales as $m_X^{-3/2}$ and is
suppressed by the large $X$ mass.

Our goal in this paper is to quantify the cross section for dark quarkonia formation.
This is, of course, a strong-coupling problem, so
we will employ two simple toy models in which the calculation is tractable.
As we will see, the results can be readily interpreted to infer the behavior of the cross section in the case of interest.

We consider a dark SU($N$) with two Dirac fermions $X$ and $q$ in the fundamental representation.
We denote the SU($N$) confinement scale by $\Lambda_D$, although much of our discussion applies to real QCD as well.
$X$ is heavy, with mass $m_{X}\gg \Lambda_D$, while $q$ is light, with $m_q\lesssim\Lambda_D$.
We denote the color-singlet heavy-light mesons by $H_X \equiv X\bar{q}$ and $\bar{H}_X \equiv \bar{X}q$.
The $X$ and $\bar{X}$, as well as their hadrons,
are stable by virtue of a flavor symmetry.

We examine two prototypical contributions to quarkonia production in $H_X$--$\bar{H}_X$ collisions.
The first is a radiation process, in which the ``brown muck'' is merely a spectator.
To isolate the contribution of the heavy $X$s, we invoke a dark U(1), 
under which $X$ is charged while $q$ is neutral.
The heavy fermions $X$ and $\bar{X}$ emit radiation in order to bind, and the relevant process is
\begin{equation}
 H_X + \bar{H}_X \to (X\bar{X}) + \varphi\qquad\text{[radiation by the $X$s]}.
\end{equation}
Here $\varphi$ is the dark photon.
Since the photon is emitted by the heavy $X$,
the cross section for this process can be calculated using non-relativistic QCD (NRQCD) with a simple potential modeling the SU($N$) interaction.
We use the Cornell potential, with a cutoff at a distance of order $1/\Lambda_D$ to simulate
the screening by the brown muck.
The resulting cross section is not geometric, but rather $m_X$-suppressed,
in accordance with the simple semiclassical estimate above.

In the second process, the brown muck plays a key role in the interaction,
leading to a geometric cross section for quarkonium formation.
This happens, for example, when the radiation is emitted by the brown muck itself, which, as a result, exerts a force on the heavy $X$.
While we cannot reliably calculate
the cross section in this case, 
we can nonetheless capture the brown muck dynamics by
considering the limit
$m_q>\Lambda_D$, in which quarkonium formation can be thought of as a rearrangement of the heavy and light quarks,
 \begin{equation}
\quad H_X + \bar{H}_X \to (X\bar{X}) + (\bar{q}q)\qquad\text{[rearrangement]}.
 \end{equation}
The cross section for this process can be  calculated
in analogy with hydrogen-antihydrogen rearrangement into protonium and positronium.
As we will see, for $m_q>\Lambda_D$,
only  the Coulombic states contribute.
The result is a geometric cross section, which scales as the square of the Bohr radius $a_q =1/(\bar{\alpha}_D m_q)$.
Thus, quarkonium production 
is effective at low temperatures not because of confinement \textit{per se}, 
but because of the large hierarchy between $a_q$ (the size of $H_X$)
and $1/m_X$ (the Compton wavelength).
We expect this result to persist  as $m_q$ is dialed back below $\Lambda_D$:
the quarkonium cross section will continue to scale as the size of $H_X$,
which, in this case, is $1/\Lambda^2_D$.

As we will see, the geometric cross section arises from summing the contributions of many large 
(i.e., $\sim a_q$-sized) \XXb bound states, for which the process is exothermic. 
These states cannot be dissociated, and will de-excite to the ground state,
in which the $X$ and $\bar X$ annihilate.
We will not discuss the cosmology of a specific model in detail, but merely sketch the essentials, following Ref.~\cite{Kang:2006yd}.
Prior to the formation of the $H_X$ and $\bar{H}_X$ mesons, the $X$ particles annihilate and freeze out in the early universe with the standard relic density
\beq
\Omega^{\ann}_X h^2 \sim \frac{10^{-9} \GeV^{-2}}{\left\langle\sigma^\ann_X v\right\rangle}\sim \left(\frac{m_X}{10^4\GeV}\right)^2 \frac{1}{\alpha^2_D(m_X)}\,.
\eeq
Following the second stage of annihilations, the $H_X$ relic abundance is given by
\beq \label{eqn:omegaf}
\Omega^f_{H_X} \sim \sqrt{\frac{\Lambda_D}{m_X}} \, \fracp{m_X}{\text{30 TeV}}{2} \,.
\eeq
Some fraction of $X$s remain in hadrons containing multiple $X$s, such as baryons.
The various final abundances are model dependent and we do not explore them in detail here.
Still, the late re-annihilations give a new mechanism for generating the relic
abundance of dark matter,
which is now a function of the two scales $m_X$ and $\Lambda_D$.
This opens up many interesting directions to explore.
We discuss some of the implications for cosmology in \secref{cosmo}.
In particular, the models can lead to a long era of matter
domination between $m_X$ and $\Lambda_D$.

Note that the $X$s can hadronize with light quarks $q$,
and the potential between them is screened at large distances.
The cosmology of these models is thus somewhat different from quirky models.
The presence of light quarks is important for yet another reason:
even if there are no photons in the theory, energy loss can proceed via the emission of light pions.
In contrast, in models with a pure SU($N$) at low energies,
the lightest particles are glueballs, whose mass is $\sim 7\Lambda_D$.

The formalism and the results in this paper can be applied more broadly. 
For example, it is applicable to any confined heavy relic---be it all the dark matter or a component thereof,
such as gluinos in split supersymmetry, messengers in gauge-mediated supersymmetry breaking, and so on.

This paper is structured as follows.
The toy model is described in Sec.~\ref{sec:model}.
The rearrangement calculation and its results are presented in Sec.~\ref{sec:rearr}. 
Section~\ref{sec:radBSF} focuses on the radiation process from the $X$s in which the brown muck is a spectator.
In Sec.~\ref{sec:cosmo} we consider the dynamics of the \XXb bound states generated by these processes and further implications for cosmology.
In the Appendix, we collect some useful results on the properties of the Cornell and linear potentials and their wavefunctions,
and discuss the details of the derivation of the cross section
used in Sec.~\ref{sec:radBSF}.


\section{Description of the Toy Model} \label{sec:model}
The minimal particle content in our models consists of two Dirac fermions, $(q,\bar{q})$ and $(X,\bar{X})$, in the fundamental
representation of a dark SU($N$).
In \secref{radBSF}, we will assume that $X$ and $\bar X$ are also charged under a U(1) gauge symmetry.
To describe the \XXb interaction, we turn to models of quarkonium~\cite{Novikov:1977dq,Eichten:1978tg}.
The Cornell potential interpolates between the Coulombic QCD potential at small
distances and the confining linear potential with string tension $\Lambda_D$ at large distances:
\beq \label{eqn:Cornellpotn}
V_{\mathrm{Cornell}}(R) = -C \frac{\alpha_D}{R}+ \Lambda^2_D R  \,,
\eeq
where $R$ is the distance between $X$ and $\bar X$, $C = (C_1 + C_2 - C_{12})/2$, and $C_{i}$ ($C_{12}$) are the quadratic Casimirs of the constituents (bound state).
Since $X$ is a fundamental of SU($N$) and we require a color-singlet bound state, $C = C_1 = C_2 = (N^2 - 1)/(2N)$.
The deep bound states of the system are then Coulombic, while the shallow states are controlled by the linear potential. 

At large distances, the attractive potential is screened by the brown muck surrounding $X$ and $\bar X$.
In QCD, for example, this distance is roughly the inverse of
the string tension $\approx400$~MeV (see, e.g., Refs.~\cite{Brambilla:2004wf, Brambilla:2010cs}).
In order to capture this screening, the potential is cut off at a distance $R_c$,%
\footnote{While this option is not pursued here, it may be interesting to use a temperature-dependent cutoff
to qualitatively capture the screening effects of the quark-gluon plasma. 
These cause large \XXb bound states to ``dissolve'' at finite
temperatures~\cite{Brambilla:2010cs,Brambilla:2011sg,Brambilla:2013dpa}.}
\beq\label{eqn:Cornellpotn_cutoff}
V(R) = \begin{cases} -\bar{\alpha}_D\left(\frac{1}{R}-\frac{1}{R_c}\right) + \Lambda^2_D \left(R-R_c\right)\,+\,V_0 & \text{for}\quad R<R_c\,,\\  V_0 & \text{for}\quad R \geq R_c\,, \end{cases}
\eeq
where  $\bar{\alpha}_D = C \alpha_D$, and $V_0$ is a constant.%
\footnote{%
  The choice of the constant $V_0$ is of course a matter of convenience, and we
  will in fact choose different constants in the rearrangement and radiation calculations.
 }
The cutoff behavior will naturally emerge in the rearrangement calculation
of \secref{rearr}, where we work in the calculable limit $m_q\gtrsim\Lambda_D$.
For $m_q \gtrsim \Lambda_D$, the attractive potential is cut off at distances
of order the Bohr radius of the heavy-light meson,%
\footnote{In this expression, $\bar{\alpha}_D$ should be evaluated at the energy scale of the inverse Bohr radius.}
\beq
a_q = \frac1{\bar{\alpha}_D m_q}\,.
\eeq
Thus, for $m_q$ sufficiently large, the problem reduces to a purely Coulombic potential,
and we can calculate the cross section in analogy with  hydrogen-antihydrogen 
rearrangement into protonium and positronium.

In \secref{radBSF} we will calculate quarkonia formation via radiation
by the $X$s. Here the linear part of the potential is important,
and the cutoff $R_c$ is introduced by hand. 
As we will describe, the choice of $R_c$ will be motivated
by a comparison with the masses of $B$ and $D$ mesons in the Standard Model.

We note that, in QCD, the string tension in the confined phase and the dimensional transmutation scale from the running of the QCD gauge coupling are approximately the same~\cite{Simolo:2006iw, Brambilla:2010cs, Patrignani:2016xqp},
whereas the deconfinement temperature is about a factor of two lower~\cite{Teper:2008yi}.
We will not be concerned with the lightest glueball state because its mass is a factor of about seven larger than the string tension~\cite{Morningstar:1999rf}.


\section{The Rearrangement Process} \label{sec:rearr}

At temperatures below $\Lambda_D$, the heavy $X$s are mostly found in $H_X$ ($X\bar{q}$) and $\bar{H}_X$ ($\bar{X}q$)  mesons. 
These mesons can further deplete through $H_X$--$\bar{H}_X$ scattering into
$(X\bar{X})$ quarkonia plus light hadrons.
For $m_q < \Lambda_D$, the calculation of the cross section for this process
requires the full machinery of perturbative NRQCD~\cite{Brambilla:2004jw} and is extremely difficult;
we will limit ourselves to the case $m_q \gtrsim \Lambda_D$.
This puts us firmly in the non-relativistic limit, in which
quarkonium production can be thought of  as rearrangement of the four partons,
\beq
H_X + \bar{H}_X \to (X\bar{X}) + (q\bar{q}).
\eeq
For $m_X\gg m_q$, the wavefunctions of the system can be calculated in the Born-Oppenheimer approximation, 
as in hydrogen-antihydrogen rearrangement into protonium and positronium~\cite{HHbarcoll, Jonsell:2001aa,HHbarBO}.
We will closely follow this calculation, applying it to the near-threshold energies of interest.

If the semiclassical arguments in Ref.~\cite{Kang:2006yd} are correct, the cross section
is expected to be geometric when the temperature is comparable to the binding energy of $H_X$,
with no $m_X$ suppression.
We verify this in the following calculation.


\subsection{Setup}
As discussed above, for $m_q$ sufficiently larger than $\Lambda_D$, only the Coulombic $(X\bar{X})$ states contribute. 
We will later comment on the validity of this approach as $m_q$ is taken below $\Lambda_D$.

\begin{figure}[t]
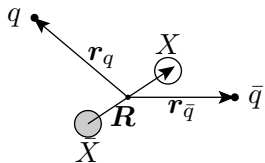

\centering
\begin{axopicture}(100,70)	
	\Vertex(15,55){1.5}
	\Text(10,55)[r]{$q$}
	\Vertex(90,25){1.5}
	\Text(95,25)[l]{$\bar{q}$}
	\ECirc(65,35){5}
	\Text(65,45){$X$}
	\GCirc(35,15){5}{0.8}
	\Text(35,5){$\bar{X}$}
	\Vertex(50,25){1}
	\Line[arrow,arrowpos=1](35,15)(65,35)
	\Line[arrow,arrowpos=1](50,25)(17,53)
	\Line[arrow,arrowpos=1](50,25)(87,25)
	\Text(48,18){$\bfR$}
	\Text(40,40){$\bfr_q$}
	\Text(70,20){$\bfr_{\bar{q}}$}
\end{axopicture}
\caption{Coordinate system used in the calculation of the rearrangement process.}
\label{fig:coordsys}
\end{figure}

The full interacting Hamiltonian of our system is written as the sum
\beq
\mathcal{H}_{\text{tot}} = \mathcal{H}_{\text{free}} + \mathcal{H}_{\text{int}} \,,
\eeq
where
\begin{align}\label{eqn:fullh}
\mathcal{H}_{\text{free}} &= -\frac{1}{m_X}\nabla^2_R-\frac{1}{2m_q}\nabla^2_{r_q}-\frac{1}{2m_q}\nabla^2_{r_{\bar{q}}} \,, \nl
\mathcal{H}_{\text{int}} &= V_{X\bar{X}}\left(R\right)+V_{q\bar{q}}\left(|\bfr_{q}-\bfr_{\bar{q}}|\right)+\mathcal{H}_{\text{tr}}\,, \\
\mathcal{H}_{\text{tr }} &=V_{q\bar{X}}\left(|\bfr_q+\bfR/2|\right)+V_{\bar{q}X}\left(|\bfr_{\bar{q}}-\bfR/2|\right)-V_{\bar{q}\bar{X}}\left(|\bfr_{\bar{q}}+\bfR/2|\right)-V_{qX}\left(|\bfr_q-\bfR/2|\right) \,. \nonumber
\end{align}
Here $\bfR$ is the vector from $\bar{X}$ to $X$ and $\bfr_q,\, \bfr_{\bar{q}}$ are the positions of $q,\, \bar{q}$ relative to the \XXb center-of-mass (CM), respectively, as shown in \figref{coordsys}. 
The potentials $V_{q\bar{X}}$, $V_{\bar{q}X}$, $V_{qX}$, $V_{\bar{q}\bar{X}}$, $V_{q\bar{q}}$, and $V_{X\bar{X}}$ are the usual Coulomb potentials (with the relevant sign for same/opposite color quarks taken into account in \eqnref{fullh}):
\beq
V(r) = - \frac{\bar{\alpha}_D}{r} \,.
\eeq 
Since we assume that \XXb are in a color-singlet configuration, this factor is the same for the six potentials.

The calculation of the rearrangement cross section involves a subtlety well known to nuclear physicists: 
the asymptotic in and out states are not eigenstates of the same free Hamiltonian, but rather eigenstates of two different interacting Hamiltonians. This is different from conventional non-relativistic scattering where $\lim_{t\rightarrow\pm\infty}\mathcal{H}_{\text{tot}}=\mathcal{H}_{\text{free}}$. In our case, the infinite past Hamiltonian is 
\begin{equation} \label{eqn:Hin}
  \mathcal{H}_{\text{in}} \equiv\lim_{t\rightarrow-\infty}\mathcal{H}_{\text{tot}} =   \mathcal{H}_{\text{free}}+V_{q\bar{X}}\left(|\bfr_q+\bfR/2|\right)+V_{\bar{q}X}\left(|\bfr_{\bar{q}}-\bfR/2|\right)\,,
\end{equation}
while the infinite future Hamiltonian is
\begin{equation} \label{eqn:Hout}
  \mathcal{H}_{\text{out}} \equiv\lim_{t\rightarrow\infty}\mathcal{H}_{\text{tot}} =   \mathcal{H}_{\text{free}}+ V_{X\bar{X}}\left(R\right)+V_{q\bar{q}}\left(|\bfr_q-\bfr_{\bar{q}}|\right) \,.
\end{equation}
The scattering cross section is then calculated in the multi-channel formalism. 
By solving the Lippmann-Schwinger equation for multi-channel scattering (see, e.g., Ref.~\cite{Taylor}), we get the simple formula for the cross section:
\beq \label{eqn:rearr_xsec}
\frac{d\sigma }{d\Omega} = (2\pi)^2 \frac{k_f}{k_i} m_Xm_q |\mathcal{M}|^2 \,,
\eeq
where $k_f$ and $k_i$ are the momenta of the final and initial states in the CM frame (see below)
and the transition matrix element is
\beq \label{eqn:rearr_amplitude}
\mathcal{M} = 2\pi \Braket{
  \Psi_f(\bfR ,\bfr_q,\bfr_{\bar{q}})| \mathcal{H}_{\text{tr}} |
  \Psi_i(\bfR ,\bfr_q,\bfr_{\bar{q}})} \,,
\eeq
where $\Psi_f,\Psi_i$ are the final- and initial-state wavefunctions and 
$\mathcal{H}_{\text{tr}} =  \mathcal{H}_{\text{tot}} -  \mathcal{H}_{\text{out}}$, as can be seen from Eqs.~(\ref{eqn:fullh}) and (\ref{eqn:Hout}). Note that in this representation of the cross section, the outgoing states $\Psi_f$ are eigenstates of $\mathcal{H}_{\text{out}}$, while the incoming states $\Psi_i$ are eigenstates of the \emph{full} Hamiltonian $\mathcal{H}_{\text{tot}}$.
Below, we discuss these states in more detail.

%

\subsection{The incoming and outgoing wavefunctions}

We wish to express our incoming and outgoing wavefunctions in the factorized form 
\beq\label{eqn:factorize}
  \Psi(\bfR ,\bfr_q,\bfr_{\bar{q}})=
  \psi^{X\bar{X}}(\bfR)\,
  \psi^{q\bar{q}}(R;\bfr_q,\bfr_{\bar{q}})
  \,.
  \eeq
In the final state this factorization is exact, since the outgoing $X$-onium and $q$-onium are asymptotically non-interacting. In other words:
\begin{equation}
  \mathcal{H}_{\text{out}} = \mathcal{H}_{X\bar{X}} + \mathcal{H}_{q\bar{q}}\, , 
\end{equation}
with
\begin{align} \label{eqn:Houts}
\mathcal{H}_{X\bar{X}} &= -\frac{1}{m_X}\nabla^2_R + V_{X\bar{X}}\left(R\right)\,,\nl
\mathcal{H}_{q\bar{q}} &= -\frac{1}{2m_q}\nabla^2_{r_q}-\frac{1}{2m_q}\nabla^2_{r_{\bar{q}}}+V_{q\bar{q}}\left(|\bfr_q-\bfr_{\bar{q}}|\right)\, .
\end{align}
The final state therefore trivially factorizes as the product of a plane wave for the outgoing $(q\bar{q})$ and 
the Coulomb bound states $\psi^{X\bar{X}}_{nlm}(R)$ (an eigenstate of $\mathcal{H}_{X\bar{X}}$) 
and $\psi^{q\bar{q}}_{100}(\bfr_q,\bfr_{\bar{q}})$ (an eigenstate of $\mathcal{H}_{q\bar{q}}$).
For concreteness, we assume that the final-state $q$-onium is in its ground state
and the $(X\bar{X})$ is static.

In contrast with the outgoing state, the incoming state is an eigenstate of the full Hamiltonian $\mathcal{H}_{\text{tot}}$, 
so we na\"ively do not expect it to factorize as in \eqnref{factorize}. 
However, we can use the Born-Oppenheimer approximation to express it in this factorized form,
\beq
\Psi_i  (\bfR ,\bfr_q,\bfr_{\bar{q}})=
\psi^{X\bar{X}}_i(\bfR)
\,\psi^{q\bar{q}}_i(R;\bfr_q,\bfr_{\bar{q}})
\,.
\eeq
Since we are in the limit $m_X\gg m_q$, this is a very good approximation:
at any given \XXb distance $R$, $q$ and $\bar{q}$ will quickly adjust their configuration, and
their wavefunction $\psi_i^{q\bar{q}}$ can therefore be calculated 
by integrating out $X$ and $\bar{X}$ and treating them as sources for the light quarks.
This gives the energy and wavefunction of the light quarks for a fixed separation $R$ between
the heavy $X$s as solutions to the eigenvalue problem
\begin{equation}
 \left[\mathcal{H}_{\text{tot}} - \mathcal{H}_{X\bar{X}}\right]\psi^{q\bar{q}}_i =
V_{\BO}(\bfR)\,\psi^{q\bar{q}}_i \,.
\end{equation}
Substituting this back into the full Schr\"odinger equation
and neglecting derivatives of $\psi^{q\bar{q}}_i$ with respect to the \XXb coordinates,
one obtains the equation for the \XXb wavefunction,
\begin{equation}
 \left[-\frac{1}{2m_X}\nabla^2_R+ V_{X\bar{X}}(R)+V_{\BO}(\bfR)\right]\psi^{X\bar{X}}_i=E_i\, \psi^{X\bar{X}}_i\,.
\end{equation}

%
\begin{figure}[ht] 
\centering
\includegraphics[scale=1]{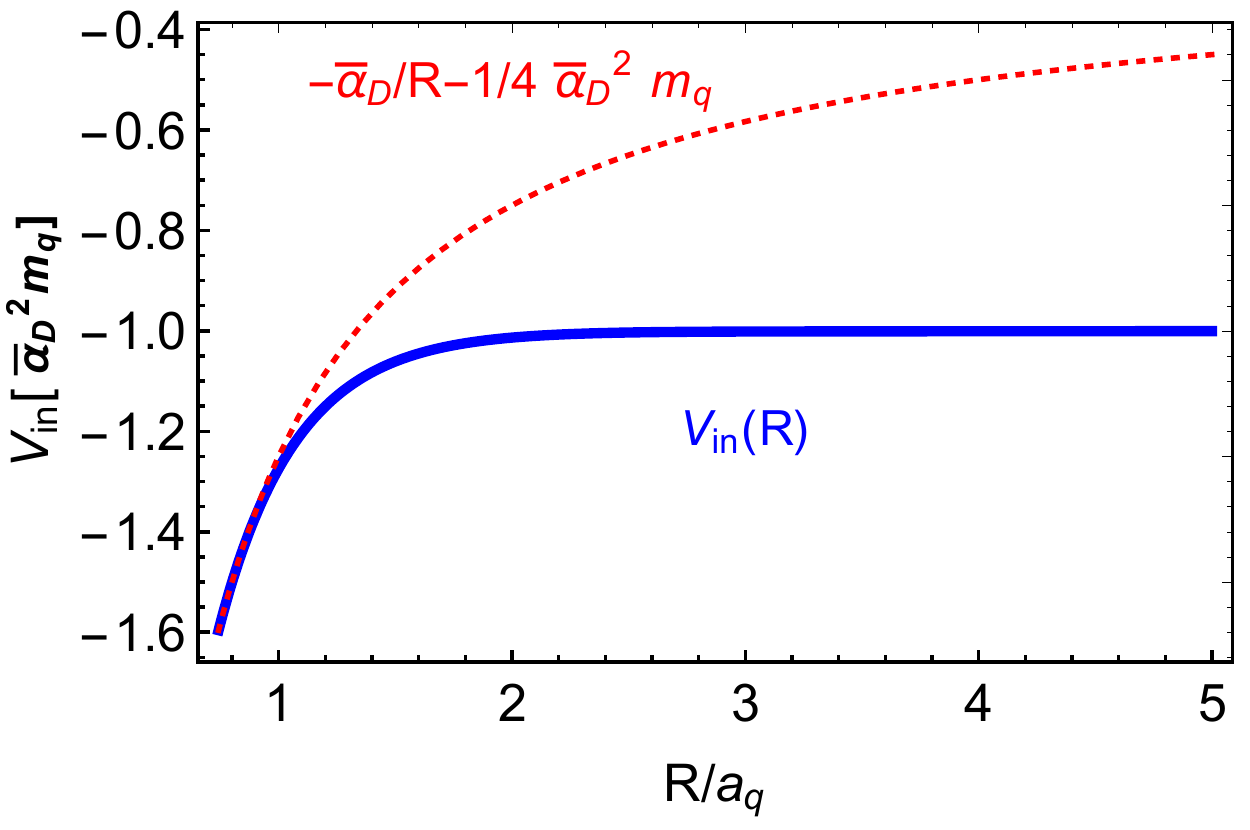}
\caption{The incoming effective potential $V_{\mathrm{in}}$ in units of $\bar{\alpha}_D^2 m_q$ for the \XXb system in the Born-Oppenheimer approximation (blue solid), as a function of \XXb separation in units of the Bohr radius $a_q = 1/(\bar{\alpha}_D m_q)$.
Also shown is the Coulomb potential for the $(X\bar{X})$ quarkonium (red dashed).
}
\label{fig:potentials}
\end{figure}
%

The effective potential for the \XXb system is then 
\beq
V_{\mathrm{in}}(\bfR)=V_{X\bar{X}}(R)+V_{\BO}(\bfR) 
\eeq
with
\beq
V_{\BO}(\bfR)=\Braket{\psi^{q\bar{q}}_i |\mathcal{H}_{\text{tot}} - \mathcal{H}_{X\bar{X}}| \psi^{q\bar{q}}_i} \,. 
\eeq
$V_{\BO}(\bfR)$ should interpolate between twice the binding energy of $H_X$,
$2E_b \equiv \bar{\alpha}_D^2 m_q$, at large $R$,
and the $q$-onium binding energy, $-\bar{\alpha}_D^2 m_q/4$, at small $R$.
Unlike in molecules, for which the Coulombic repulsion of the nuclei must be overcome,
here the two heavy particles attract each other, so we do not expect a significant potential barrier.
These na\"ive expectations are borne out in the calculation of Ref.~\cite{BOpotn}.
Since  $V_{\BO}(\bfR)$ does not depend on the initial energy of the system or on the mass $m_X$, 
we can extract $V_{\BO}(\bfR)$ from Ref.~\cite{BOpotn}.
We plot $V_{\mathrm{in}}$ in Fig.~\ref{fig:potentials}:
as expected, the effects of the light quarks captured in $V_{\BO}(\bfR)$ set in for $R$ of order $a_q$. 
Their main effect is to screen the \XXb interaction at large $R$;
in practice, this happens for $R\sim 2a_q$.

\begin{figure}[ht] 
\centering
\includegraphics[width=0.8\textwidth]{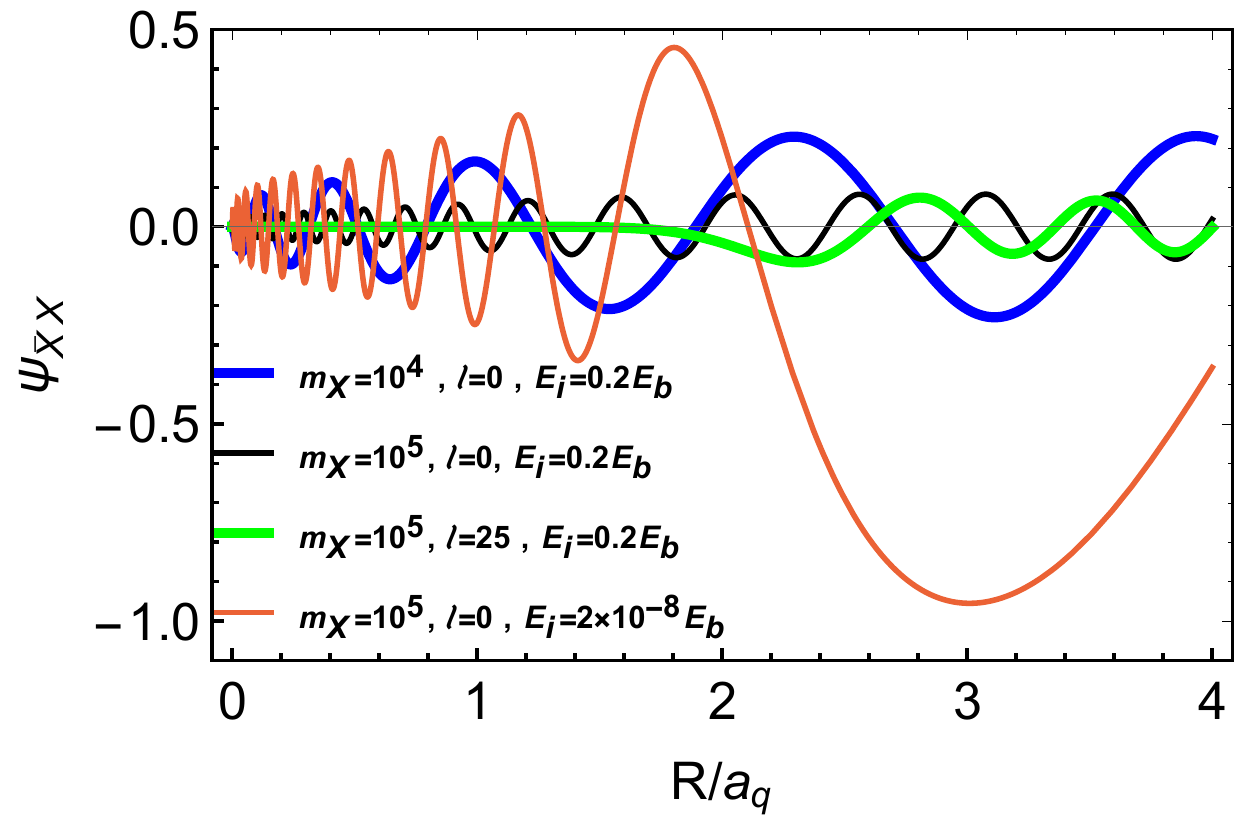}
\includegraphics[width=0.8\textwidth]{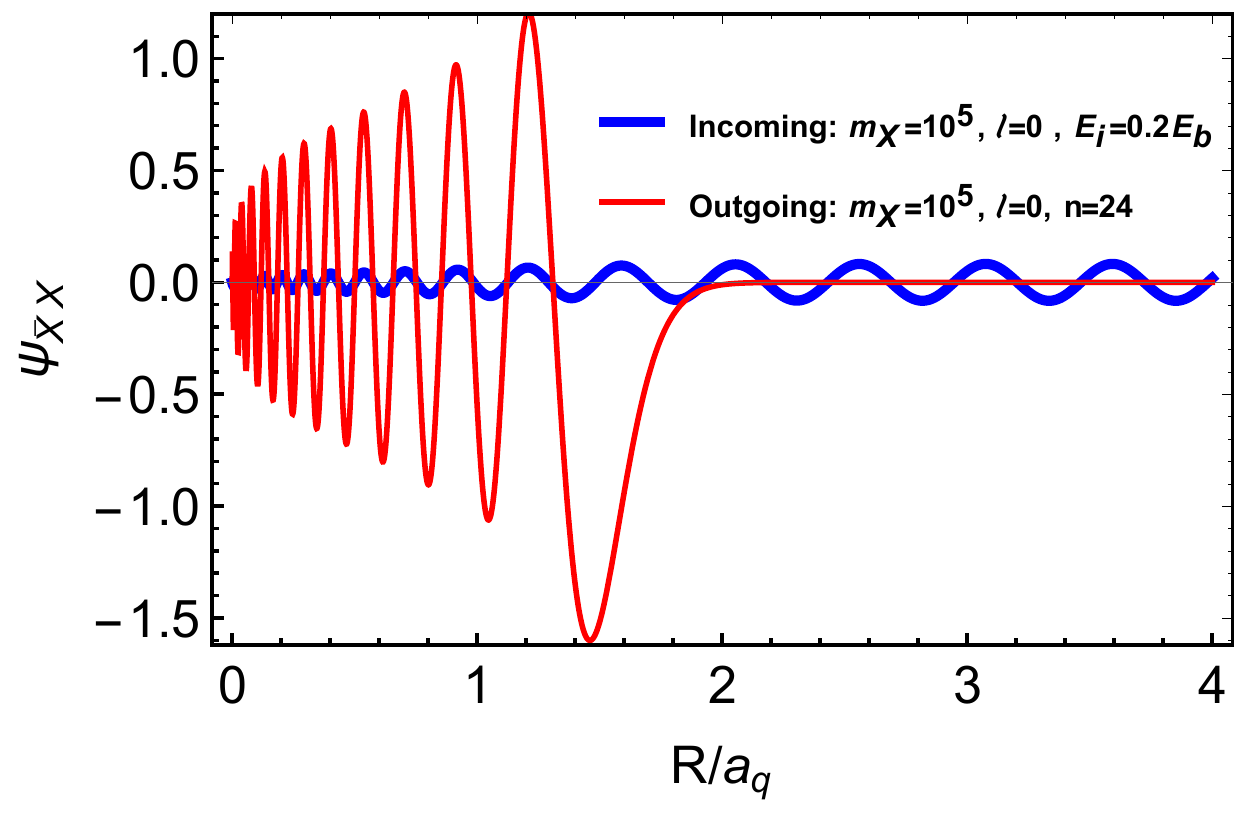}
\caption{Examples of incoming wavefunctions with various $m_X, E_i, l$ (top) and a comparison of an incoming and outgoing wavefunction (bottom). 
  $m_X$ is given in units of the inverse Bohr radius $\bar{\alpha}_D m_q$ and
  $E_b = \frac{1}{2}\bar{\alpha}_D^2 m_q$ is the $H_X$ binding energy.
  }
  \label{fig:inc_out_wavefunctions}
\end{figure}

Since $V_{\mathrm{in}}(\bfR)$ approaches a constant at large $R$, the \XXb wavefunction at large distances ($R\geq 4a_q$) is the standard free-particle solution,
\beq\label{eqn:pw}
\Psi_i(\bfR ,\bfr_q,\bfr_{\bar{q}}) \xrightarrow{R \geq 4a_q} \sum_l i^l\sqrt{2l+1}\, e^{i\delta_l} \left[\cos \delta_l \,  j_l(k_i R) -\sin \delta_l \, n_l(k_i R)\right]  Y_{l0}(\theta_R)\psi^{q\bar{q}}_i(\bfR;\bfr_q,\bfr_{\bar{q}})\,. 
\eeq
The wavefunctions for $R \leq 4a_q$ are found numerically, while their normalization is fixed by matching to \eqnref{pw} at $R=4a_q$.
Some examples for the incoming and outgoing wavefunctions are given in Fig.~\ref{fig:inc_out_wavefunctions}.

%
%


\subsection{The matrix element for rearrangement}\label{sec:me}
Using the factorized incoming and outgoing wavefunctions, the transition matrix element \eqnref{rearr_amplitude}
can be written in position space as
\beq
\mathcal{M} = \int d^3 \bfR\,\psi^{X\bar{X}*}_f(\bfR)\, \psi^{X\bar{X}}_i(\bfR)\,T(\bfR)\, ,
\eeq
where
\beq \label{eqn:T(R)}
T(\bfR) = \int d^3 \bfr_q\,d^3 \bfr_{\bar{q}}\,\,
\psi^{q\bar{q}*}_f(\bfr_q,\bfr_{\bar{q}})\,
\mathcal{H}_{\text{tr}} \,
\psi^{q\bar{q}}_i(R;\bfr_q,\bfr_{\bar{q}})\, .
\eeq
We will assume that the angular part of $T(\bfR)$ is  constant. 
This is justified when the $(q\bar{q})$ is in the ground state, 
and in the short-distance approximation for the plane wave of the $(q\bar{q})$ relative to the $(X\bar{X})$. 
The second condition is broken when $k^2_f$ becomes large enough, where we expect an $\Order(1)$ correction. 
We neglect this correction in this work, since we are mostly interested in the parametric behavior of this process. 

It is easy to see that $T(R)$ is appreciable only for $R$ of order $a_q$.
For $R\gg a_q$, this can be seen by substituting
$\mathcal{H}_{\text{tr}}=  V_{\BO}(\bfR) - \mathcal{H}_{q\bar{q}}$ in \eqnref{T(R)}.
$\psi^{q\bar{q}}_f$ is an eigenfunction of $\mathcal{H}_{q\bar{q}}$, so one is left with the overlap integral of
$\psi^{q\bar{q}}_f$ and the asymptotic $\psi^{q\bar{q}}_i$ at large $R$.
These solve the same Hamiltonian with different eigenvalues:
the former is a bound state and the latter is a continuum state.
For small $R$, the initial- and final-state wavefunctions in the integrand do overlap, but $\mathcal{H}_{\text{tr}}$ tends to zero. 

The full calculation of $T(R)$ is complicated.
In fact, the Born-Oppenheimer approximation breaks down for $R \lesssim 0.74 a_q$,
where the $q$ and $\bar{q}$ are no longer bound to their respective $X$~(see, e.g., Ref.~\cite{HHbarBO}). 
Still, we can use this approximation to get a rough estimate of the cross section.
In particular, $T(R)$ is independent of the mass $m_X$ in the Born-Oppenheimer approximation, 
so we can extract $T(R)$ from Ref.~\cite{HHbarBO}.%
\footnote{The $m_X$ dependence will enter through higher-order corrections in the effective theory, 
and will be suppressed by some fractional power of $m_X$. Here we are only interested in the leading result.}
The result in the relevant range ($R\sim a_q$) can be parametrized as
\begin{equation}
 T(R) =
\begin{cases}
 \beta \left[E_f +\frac{1}{4}\bar{\alpha}_D^2 m_q - V_{\BO}(R)\right] & \text{for}\quad R>  0.74a_q\,,\\
0 &\text{for}\quad R \le  0.74 a_q\,,
\end{cases}
\end{equation}
where $E_f$ is the kinetic energy in the final state and $\beta$ is an $\Order(1)$ factor determined by matching to the hydrogen-antihydrogen results.
Evidently, $T(R)$ depends on the binding energy of the
$(X\bar{X})$ quarkonium,
$E_b^{X\bar{X}}$, since
$E_f = E_i + E_b^{X\bar{X}} - \frac34\bar{\alpha}_D^2 m_q$.


\subsection{Rearrangement results}
\begin{figure}[ht!] 
\centering
\includegraphics[width=0.79\textwidth]{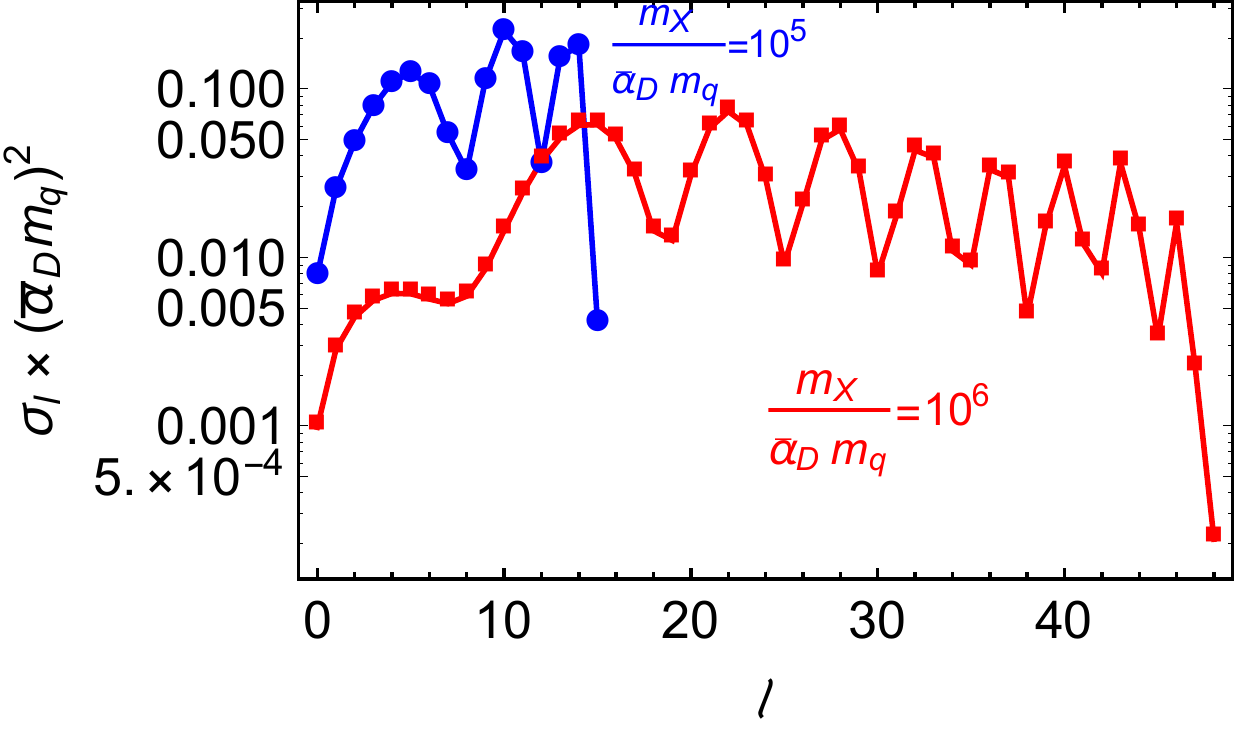}
\caption{The rearrangement cross section for each partial wave $l$,
for two different values of $m_X$, with $E_i = 0.6 E_b$.
All partial waves in the incoming state contribute up to a maximal $l\equiv l_{\max} \sim  k_i a_q$.
}
\label{fig:rearr_results_NL}
\end{figure}

We calculate the cross section of \eqnref{rearr_xsec} for different masses $m_X$ and incoming kinetic energies $E_i$, keeping $a_q$ fixed. 
In the approximation we use (see \secref{me}), 
the angular part of the overlap integral is translated into a selection rule $l_{X\bar{X}} = l_i \equiv l$.
The breakdown of the cross section into partial waves---or $(X\bar{X})$ angular momenta---is given in
Fig.~\ref{fig:rearr_results_NL} for high incoming kinetic energy $E_i = 0.6 E_b$ and $\bar{\alpha}_D = 1/137$.
We see that it vanishes above some maximal $l$,  which corresponds to the classical angular momentum
\beq
l_\max \sim k_i a_q = \sqrt{E_i m_X} a_q \sim  \sqrt{E_i m_X}/m_q\,.
\eeq

%
%

\begin{figure}[ht] 
\centering
\includegraphics[width=0.49\textwidth]{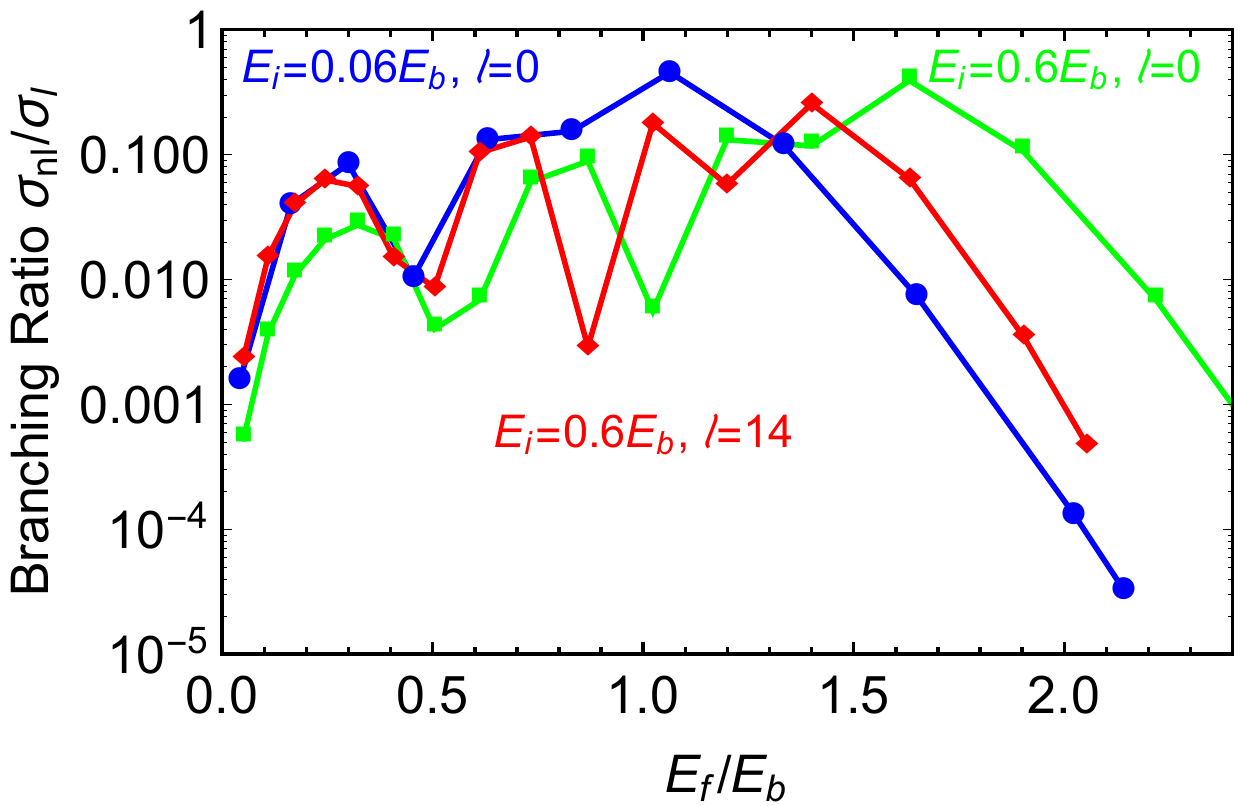}
\includegraphics[width=0.49\textwidth]{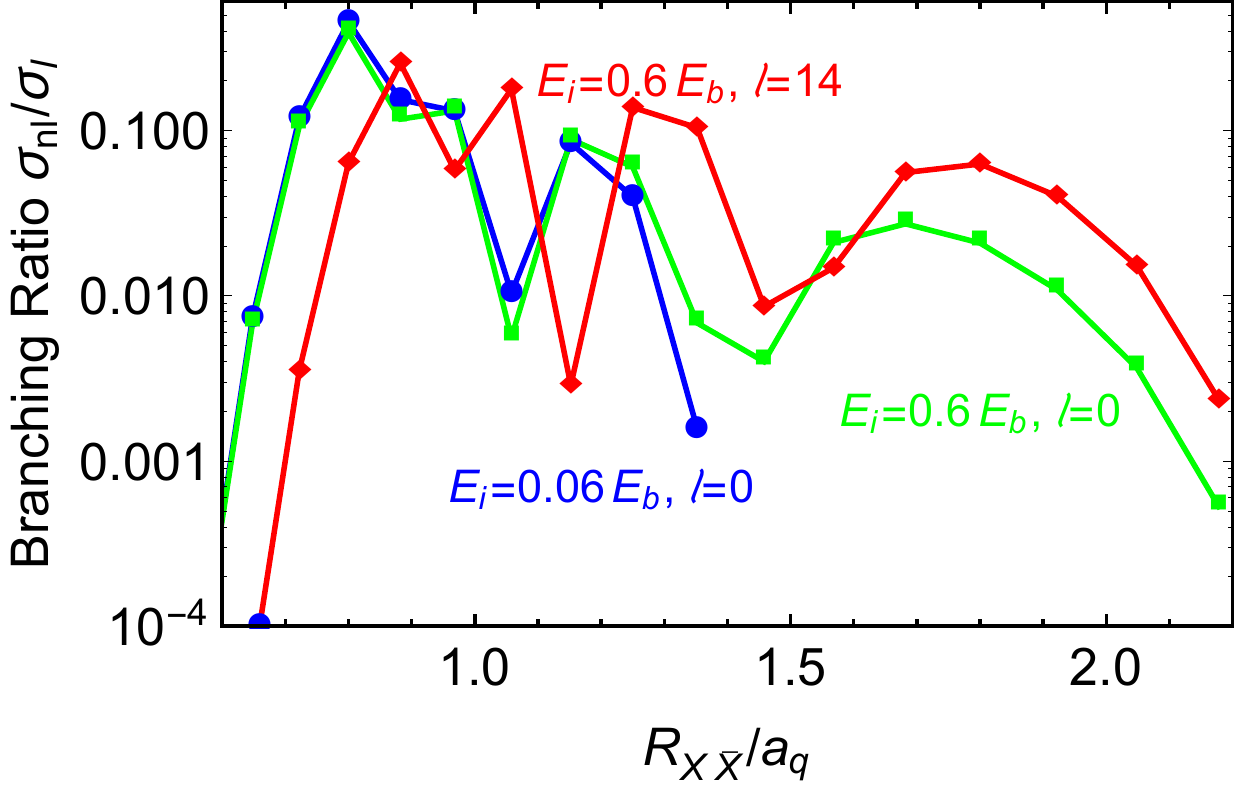}
\caption{The branching fraction $\sigma_{nl}/\sigma_l$ for some initial partial waves to form an
$(X\bar{X})$
quarkonium
of definite $n,l$ (uniquely defined by the $x$-axis).
The results are presented for $m_X = 3\times 10^4 \bar{\alpha}_D m_q$ and several values of $E_i$ and $l$. 
Left panel: 
The branching fraction as a function of the kinetic energy in the final state.
Right panel: 
The branching fraction as a function of $R_{X\bar{X}}$, the mean radius of the final state.
}
\label{fig:rearr_results_SIZE}
\end{figure}

It is also interesting to examine the branching fraction $\sigma_{nl}/\sigma_l$ for some initial partial wave to form an
$(X\bar{X})$ quarkonium of definite binding energy.
In Fig.~\ref{fig:rearr_results_SIZE} (left panel), we show this branching fraction for $l = 0, 14$,
as a function of the final kinetic energy for two values of the initial kinetic energy $E_i/E_b = 0.6, 0.06$.

%

%
\begin{figure}[ht] 
\centering
\includegraphics[width=0.7\textwidth]{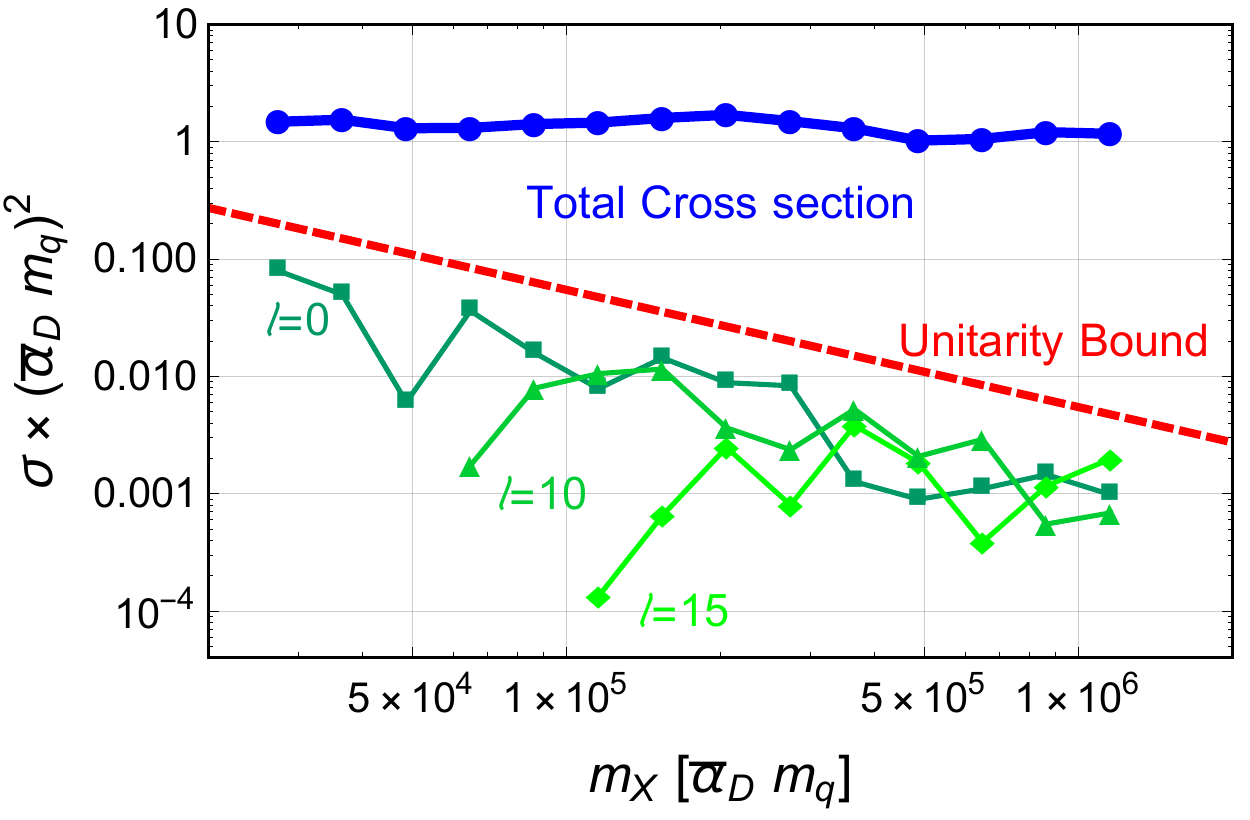}
\caption{The rearrangement cross section. 
The blue line is the total cross section for an incoming energy $E_i = 0.6 E_b$,
and is geometric.
Several individual partial-wave contributions ${\sigma_l}/{(2l+1)}$ are given in green, 
together with the unitarity bound $4\pi/k^2_i$ (red dashed line) for $l < l_\max \sim k_i a_q$.
\label{fig:rearr_results}}
\end{figure}
%

We see that quarkonium formation by rearrangement is an exothermic process:
the kinetic energy in the final state does not vanish even when the initial momentum is taken to zero.
Therefore the inverse process shuts off at low energies $\lesssim E_b$. 
Furthermore, only quarkonia with binding energies around $E_b$ are produced at $T\sim E_b$.
The cross sections drop to zero for large final-state energies corresponding to
$(X\bar{X}$) binding energies above $E_b$. 
Thus, deep Coulombic  bound states with binding energies
$E_b^{X\bar{X}} \sim \bar{\alpha}_D^2 \, m_X$ are not formed in the rearrangement process.
Correspondingly, the bound states produced are large, with size $\sim a_q$.
This behavior is clearly exhibited in Fig.~\ref{fig:rearr_results_SIZE} (right panel),
where we show the cross section as a function of the mean quarkonium radius $R_{X\bar{X}}$.

The total cross section for an initial energy of order the $H_X$ binding energy is plotted in Fig.~\ref{fig:rearr_results} (blue line) as a function
of $m_X$. 
Indeed, the cross section is geometric, $\sigma \sim a_q^2$, and is independent of $m_X$ to a very good approximation.
It is interesting to compare the partial-wave contributions with the unitarity bound,
\beq \label{eq:partial_waves_xsec}
\sigma_l \leq  (2l+1)\,\frac{4\pi}{k^2_i} \,.
\eeq
We therefore also plot in this figure several individual partial-wave contributions
normalized by $2l+1$ (green lines), 
compared to $4\pi/k^2_i$ (red dashed line). 
Clearly, for initial-state energies close to the binding energy of $H_X$,
the cross section for each partial wave lies close to the unitarity bound.
Summing over all the partial waves up to $l_\max$,
\beq
\sigma = \sum_{l=1}^{l_\max} \sigma_l\sim \frac{4\pi}{k^2_i} \sum_{l=1}^{l_\max} (2l+1)\sim
\frac{4\pi}{k^2_i} l^2_\max \sim 4\pi a_q^2 \,.
\eeq
Thus the total cross section is geometric, and scales with the area of the $H_X$ bound state. 
In the low-temperature limit, we have checked that
only $s$-wave processes are non-vanishing and the cross section scales as
\beq
\sigma \sim \frac{1}{k_i} a_q \,,
\eeq
as demonstrated in Refs.~\cite{HHbarcoll, HHbarBO}.
 
The above results apply to pure U(1) models.
In the context of an SU($N$), we have explicitly seen that for $m_q$ above $\Lambda_D$, 
the light quarks truncate the \XXb attraction for $R \gtrsim a_q$ via $V_\BO$, long before the linear potential sets in.
This justifies neglecting the linear potential in the rearrangement calculation.

We can now turn to the limit of interest, $m_q$ below $\Lambda_D$.
As we have seen above, the cross section scales with 
the size of the $H_X$ bound state, $a_q \gg 1/m_X$,
thanks to the large number of partial waves contributing.
This behavior is not special to the purely Coulombic case.
In fact, the Coulombic contribution gives a conservative estimate
of the cross section generated by the Cornell potential.
Thus we expect a geometric cross section
for $m_q$ below $\Lambda_D$, with the Bohr radius $a_q$
replaced by $1/\Lambda_D$.

The \XXb bound states produced via rearrangement at $T\sim E_b$
are of size $\sim a_q$, much larger than the Compton wavelength of $X$. 
However, since the process is exothermic, these \XXb bound states cannot be dissociated at $T \lesssim E_b$. 
In the case at hand, since there are light pions (or mesons) in the theory,
nothing impedes the relaxation of these states to the ground state,
in which the \XXb pair annihilates.


\section{The Radiation Process: Spectator Brown Muck} \label{sec:radBSF}

In the rearrangement process described above, the brown muck plays a central role.
It is instructive to contrast this with a process in which the brown muck is merely a spectator.
As we will see, in this case, the cross section scales with $m_X$.

To isolate the dynamics of the heavy quarks, we take  $X$ to be charged under a dark U(1), while the light quarks are neutral.
The relevant bound states are large, of size $\sim 1/\Lambda_D$, and are described by the linear part of the Cornell potential.
To calculate $(X\bar{X})$ quarkonium production
for $T\lesssim\Lambda_D$, we can therefore neglect the Coulombic part of the potential.
This is also consistent with previous studies showing that, for pure U(1) models with no light charged particles,
$(X\bar{X})$ bound-state formation gives only mild modifications of the $X$ relic abundance~\cite{Feng:2009mn, vonHarling:2014kha}.
For bound-state formation to deplete the $X$ abundance by orders of magnitude, a new scale is required.
In the radiation process considered here, the new scale is $\Lambda_D$.

The radiative quarkonium production process is then 
\beq
H_X + \bar{H}_X \to (X\bar{X}) + \varphi \,,
\eeq
where $\varphi$ is a (dark) photon that couples only to $X$;
however, our results below apply more generally to other light particles which can be emitted by $X$. 
Unlike in the previous section, here the light quarks $q$ are relativistic.

We will follow the field-theoretic formalism for computing bound-state
formation cross sections with long-range interactions detailed in Ref.~\cite{Petraki:2015hla}.
Alternatively, these results can be obtained using the standard non-relativistic QM approach for calculating transition amplitudes, 
treating the photon as a classical field~\cite{Wise:2014jva, An:2016gad}.

The first step is to derive the spectrum and two-particle wavefunctions that describe
the bound and scattering states of the \XXb system.
While the light quarks do not actively participate in the radiation process,
they screen the heavy $X$s at large distances.
This is captured by the cutoff $\rc$ in the potential of \eqnref{Cornellpotn_cutoff},
and leads to a continuum of $H_X$--$\bar{H}_X$ states with energies above the
open $H_X$--$\bar{H}_X$-production  threshold.
Roughly speaking, the hadron mass is given by the sum of the heavy constituent masses,
with each light quark or gluon contributing about $\Lambda_D$ to the mass.
More precisely,
\beq
m_{H_X}= m_X + \kL\Lambda_D+ \Order(\Lambda^2_D/m_X^2) \,,
\eeq
where $\kL$ is an $\Order(1)$ constant~\cite{Manohar:2000}. 
The spectrum of bottom and charm mesons in QCD suggests $\kL\Lambda_D \sim 600$~MeV, with $\Lambda_D \sim 400$~MeV, 
so in the following, we set $\kL=1.5$.
To estimate the cutoff $\rc$, we use the fact that the maximal $H_X$ binding energy, $E_b^\max$,
coincides with the onset of the continuum,
\beq \label{eqn:Ecutoff}
2m_X + E_b^\max = 2m_{H_X}\,. 
\eeq
Since the maximal bound state energy of the linear potential is
\beq
E_b^\max =\Lambda^2_D\rc+\frac{l(l+1)}{m_X\rc^2}\sim \Lambda^2_D \rc\,,
\eeq
we set the cutoff to
\beq \label{eqn:rcutoff}
\rc = \frac{2\kL}{\Lambda_D}= \frac{3}{\Lambda_D} \,.
\eeq

In summary, the potential we consider is
\beq \label{eqn:lin_cutoff}
V(R) = \begin{cases}
\Lambda^2_D \left(R-R_c\right) & \text{for}\quad R<R_c\,, \\  0 &\text{for} \quad R \geq R_c\,, \end{cases}
\eeq
with $\rc$ given by \eqnref{rcutoff}
and $V_0$ of \eqnref{Cornellpotn_cutoff} chosen as zero for convenience. 
Defining
\beq \label{eqn:R0E0}
R_0 \equiv \left(\frac{\Lambda_D}{m_X}\right)^{1/3} \frac{1}{\Lambda_D}\,,\qquad
E_0 \equiv \left(\frac{\Lambda_D}{m_X}\right)^{1/3} \Lambda_D \,,
\eeq
which are the characteristic splittings in energy and size between
successive states, the radial part of the wavefunction,  $\chi_{ln}$, solves
\beq \label{eqn:radialred}
-\chi_{ln}''(x) + V^l_{\text{eff}}(x) \, \chi_{ln}(x) = \epsilon_{ln}  \chi_{ln}(x) \,, 
\eeq
where $x = R/R_0$, $\epsilon_{ln}= {E_{ln}}/{E_0}$, and
\beq
V^l_\eff(x) = \frac{l(l+1)}{x^2} + x - x_c
\eeq
with $x_c = {\rc}/{R_0}$.

Using the semiclassical approximation, we can estimate the maximal angular momentum $l\w{max}$ of the bound states and the energy $\epsilon\w{min}$ of the lowest bound state with a given $l$.
The lowest energy bound state for each $l$ classically corresponds to a minimum of the effective potential; the position $x^l\w{min}$ and the energy $V^l\w{eff}(x^l\w{min})$ of the minimum must satisfy $x^l\w{min}<x_c$ and $V^l\w{eff}(x^l\w{min})<0$, which result in
\begin{equation}\label{eqn:lmax}
 l\w{max}\simeq\sqrt{\frac{4x_c^3}{27}}\,,
\qquad
 \epsilon\w{min}\simeq3\fracp{l}{2}{2/3}-x_c\,.
\end{equation}
In \appref{eigenCornell}, we collect some results for the effective potential and radial wavefunctions  for various choices of the parameters.


\subsection{Radiation results}

\begin{figure}[t]
\centering
 \includegraphics[scale=0.72]{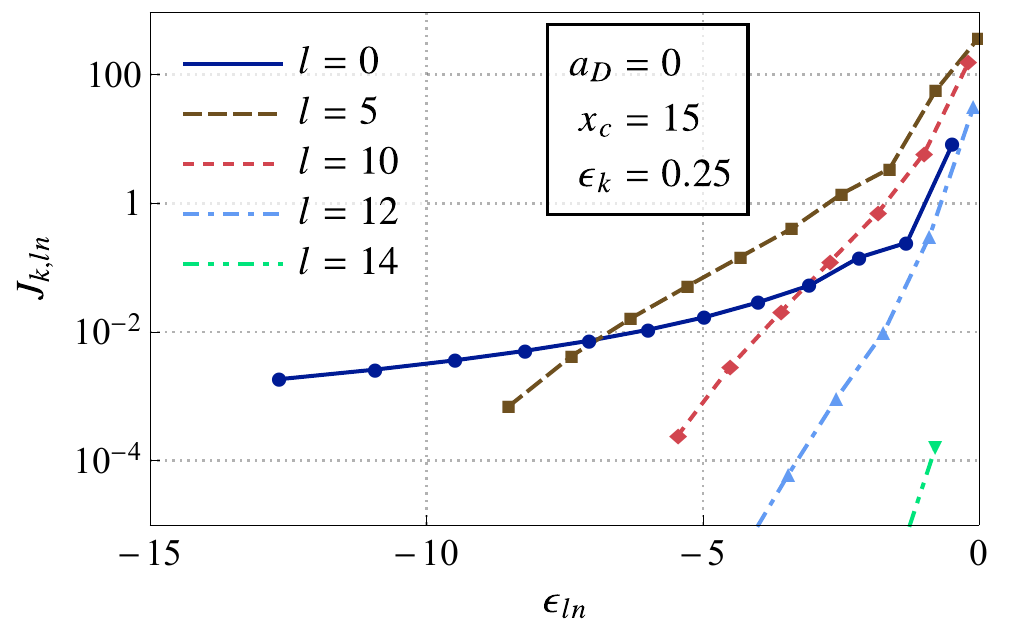}
\caption{
$J_{k,ln}$ as a function of $\epsilon_{ln}$ for $\xc=15$ and $\epsilon_k=0.25$.
The lines correspond to different values of $l$.
}
\label{fig:J}
\end{figure}

The cross section for $H_X + \bar{H}_X \to (X\bar{X})_{lmn} + \varphi$ in the CM frame of the initial state is given by
\beq \label{eqn:radxs}
v_{\text{rel}} \frac{d\sigma_{\bfk \to lmn}}{d\Omega} = \frac{|\bfPphi|}{128 \pi^2 m^3_X}\left| \bm{\mathcal{M}}_{\bfk \to lmn} \right|^2 \, ,
\eeq
where $\bfPphi$ is the three-momentum of the radiated light state and, assuming that $\varphi$ is massless,
\beq
|\bfPphi| = E_k-E_{ln} \,,
\eeq
where $E_k$ is the kinetic energy of the initial state.
We calculate this cross section in \appref{matelem}.
It is useful to write the cross section in terms of dimensionless quantities as
\beq \label{eqn:radxsdimless}
v_{\text{rel}}\, \sigma_{k\hat{\bfz} \to ln} =
\sum_{m=-l}^{l}  v_{\text{rel}}\,\sigma_{k\hat{\bfz} \to lmn} =
\frac{2 e_X^2}{m_X^2} \left(\frac{\Lambda_D}{m_X}\right)^{2/3}J_{k,ln}\,,
\eeq
where $e_X$ is the U(1) charge of $X$, $\e_k = {E_k}/{E_0}$,
\beq \label{eqn:J}
J_{k,ln}=(\epsilon_k-\epsilon_{ln})^3 \left[ (l+1)\left|I_{k,l+1 \to ln}\right|^2
+ l\left|I_{k,l-1 \to ln}\right|^2 \right] \,,
\eeq
and $I$ is the radial wavefunction overlap integral
\beq
I_{k,l\pm 1 \to ln} = \int dx \, x \, \chi^*_{ln}(x)\, \chi_{k,l\pm1}(x) \,.
\eeq
We plot $J_{k,ln}$ for several values of $l$ in Fig.~\ref{fig:J}, fixing $m_X/\Lambda_D = 125$ 
and the initial kinetic energy $E_k/\Lambda_D = 0.05$.
As expected, the large, shallow bound states give the largest
contributions.

%
%

\begin{figure}[t] 
\centering
\includegraphics[scale=0.72]{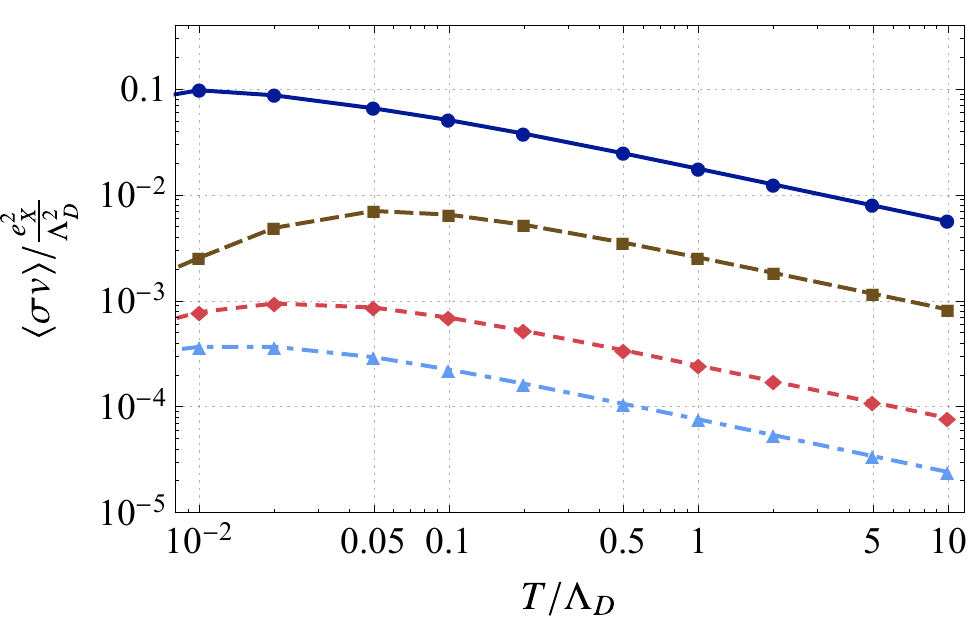}
\hfil
\includegraphics[scale=0.72]{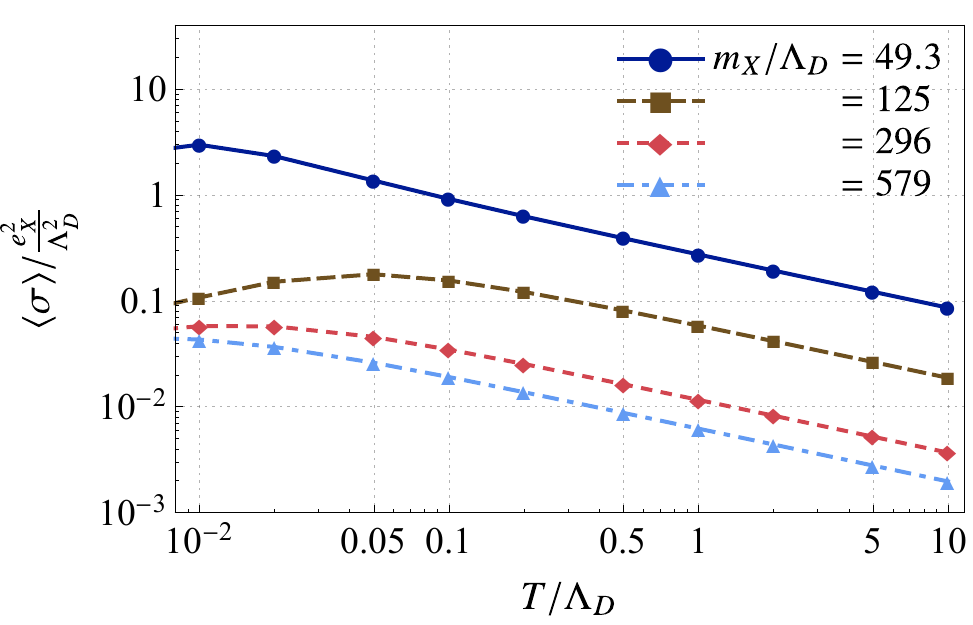}
\caption{Results of thermally-averaged radiative
quarkonium production cross section $\mean{\sigma v}$ (left panel) and $\mean{\sigma}$ (right panel) as a function of temperature for different values of $m_X/\Lambda_D$.
}
\label{fig:xc_rad_results}
\end{figure}

The total thermally-averaged cross section $\mean{v\w{rel}\sigma}$ (see \appref{thermalxs}) 
is shown as a function of the temperature in \figref{xc_rad_results} (left panel) for several choices of $m_X$.
We also show $\mean{\sigma}$ for the same parameters (right panel).
The cross section is clearly dependent on $m_X$, and decreases as $X$ becomes heavier.
In fact, for $T\gtrsim 0.1\Lambda_D$, the scaling is well described by $\mean{v\w{rel}\sigma}\propto m_X^{-2}$ and $\mean{\sigma}\propto(m_X^3\Lambda_D)^{-1/2}$, which agrees with the semiclassical estimate in \secref{intro}.
Thus, for the high mass region of interest, $m_X\gg \Lambda_D$, this contribution is negligible
compared to processes mediated by the brown muck, such as the rearrangement process.
We expect this qualitative behavior to persist regardless of the spin of the radiated particles.

%
%

\section{Implications for Cosmology} \label{sec:cosmo}
We have found that, at temperatures below the confinement scale $\Lambda_D$, 
$(X\bar{X})$ bound states are formed with a geometric cross section with no $m_X$ suppression.
These bound states are of size $\sim 1/\Lambda_D$, but since the process is exothermic, they cannot be dissociated, 
and eventually de-excite to the ground state, in which the \XXb pair annihilates.
The rate for this de-excitation process depends mainly on the light degrees of freedom.
For the large $(X\bar{X})$ bound states produced, the level splittings are of order $(\Lambda_D/m_X)^{1/3}\Lambda_D$, 
so we need massless photons or light pions in order to have allowed transitions (as in the models we considered here). 
In the case where the DM is charged under real QCD, the rate should be sizable,
so we expect any $(X\bar{X})$ bound states to quickly decay to light particles.%
\footnote{In models with no light degrees of freedom, as in the case of an adjoint $X$ with no light quarks~\cite{Feng:2011ik},
the lightest degrees of freedom are glueballs of mass $\sim 7\Lambda_D$, and this rate is suppressed by powers of $\bar{\alpha}_D^l$.}

As a result, the abundance of $H_X$ hadrons is depleted by this second stage of annihilations, down to~\cite{Kang:2006yd}
\beq \label{eqn:omegaf2}
\Omega^f_{H_X} \sim \sqrt{\frac{\Lambda_D}{m_X}} \fracp{m_X}{\text{30 TeV}}{2} \,,
\eeq 
which can be much smaller than the relic density from perturbative \XXb annihilations earlier in the thermal history. 

There are, however, various different hadronic states (in addition to $H_X$) in which $X$ can
survive~\cite{Jacoby:2007nw}.
These include other types of single-$X$ hadrons such as baryonic
$Xq^{N-1}$ hadrons, double-$X$ states such as baryonic $XXq^{N-2}$,
and up to purely heavy baryons $X^N$. 
Thus, in general, our simple toy model can produce multi-component DM with different masses and relic abundances.

We leave a detailed investigation of the parameter space of the models for future study, but note a few qualitative features.\footnote{Many of the relevant
processes were described in Ref.~\cite{Jacoby:2007nw} for TeV-mass colored relics.}
The cross section for producing double-$X$ states should be comparable to the
quarkonium cross section we calculated,
albeit smaller by $\Order(1)$ factors because of the smaller binding energies in this case.
However, for $N>2$, the double-$X$ hadrons contain some light quark(s), so their size is $1/\Lambda_D$.
They can therefore interact quite efficiently with hadrons containing $\bar X$ to form 
bound states containing both $X$ and $\bar{X}$, where \XXb pairs can quickly annihilate.
Meanwhile, they can also interact with $H_X$ hadrons and form triple-$X$ hadrons, 
and this chain may go on until pure-$X$ (or pure-$\bar{X}$) baryons are formed.
All these processes should have geometric cross sections.
The pure-$X$ baryons eventually de-excite to the ground state, the size of which is much smaller than $1/\Lambda_D$. 
Their interactions with other hadrons therefore shut off.
As a result, the $X$s inside the baryons remain as stable relics, while the other $X$s may effectively be annihilated.
A systematic analysis necessitates solving the Boltzmann equation with these dynamics, which is beyond the scope of this work.
However, it is safe to assume that a sizable fraction of the original $X$s can be preserved in a variety of stable hadronic relics.

For DM charged under real QCD, heavy-light hadrons, with size $1/\Lambda_D$, are subject to stringent constraints, 
but they are efficiently depleted at $T\lesssim\Lambda_D$ as we have shown.
In contrast, the relic abundance of $X^3$ baryons may be just somewhat smaller than the original $X$ relic abundance.
Thus the fundamental $X$s we considered here may give a different realization of colored DM~\cite{DeLuca:2018mzn}, 
but whether the models are viable requires a more detailed analysis. 
Note that the scenario considered in Ref.~\cite{DeLuca:2018mzn},
namely a heavy Dirac adjoint $X$, is non-generic, in that two $X$s can form a
stable singlet with no additional light quarks.

In summary, 
the simple models considered here typically  give rise to several components of DM composites of different masses.
For certain choices of $m_X$ and $\Lambda_D$, these can exhibit self-interactions,
transitions between different excited states, and, depending on the coupling to the Standard Model, 
modified direct and indirect detection cross sections.
In some variants of these models,  the DM abundance can be significantly
depleted at $\Lambda_D$, leading to a long era of matter domination between $m_X$ and $\Lambda_D$.


\section{Conclusions} \label{sec:conc}

In this paper we have considered the cosmological dynamics of bound states that are much larger than their inverse mass, 
taking as an example  $X\bar{q}$ mesons in a confining theory where $X$
is much heavier than $q$ and the confinement scale.
We calculated the cross section for quarkonium production from
heavy-light meson scattering.
The cross section is geometric, and scales with the area of the incident
heavy-light mesons.
The relic density of heavy-light $X$-mesons is therefore efficiently
diluted with rates much higher than the $s$-wave unitarity bound due to the
participation of many partial waves in the process. 
We also find that the process is mainly mediated by the effective interaction of the light quarks.
In contrast, processes in which only the heavy constituents participate
have mass-suppressed rates.
It is amusing to note that if the lifetime of $B$-mesons were longer,
such processes could be experimentally measured.


\section*{Acknowledgments}
The authors thank Yuval Grossman,  Marek Karliner, Teppei Kitahara,  Peter Lepage, Michael Peskin,
 Jon Rosner, Ben Svetitsky, John Terning,
and especially Markus Luty for useful discussions.
OT thanks Barak Hirshberg for his insight in molecular quantum mechanics.
The authors are grateful to the Mainz Institute for Theoretical Physics (MITP) and to the Aspen Center for Physics, supported in part by NSF-PHY-1607611, for hospitality and partial support during the completion of this work.
This work is supported
by the Israel Science Foundation (Grant No.~720/15),
by the United-States--Israel Binational Science Foundation (BSF) (Grant No.~2014397),
by the ICORE Program of the Israel Planning and Budgeting Committee (Grant No.~1937/12).
MG is supported by the NSF grant PHY-1620074 and the Maryland Center for Fundamental Physics.
SI is supported at the Technion by a fellowship from the Lady Davis Foundation and at Padova University by the MIUR-PRIN project 2015P5SBHT 003 ``Search for the Fundamental Laws and Constituents''.
GL acknowledges support by the Samsung Science \& Technology Foundation.
OT is supported in part by the NSF grant PHY-1719877.

\appendix


\section{Cross Section for Bound-State Formation in the Radiation Process} \label{app:matelem}

In this section we summarize the calculation of the bound-state formation cross section via radiation discussed in \secref{radBSF}.
We also collect here some useful results on the spectrum and wavefunctions of the Cornell and linear potential.


\subsection{Eigenstates of the Cornell potential} \label{app:eigenCornell}

We model the \XXb attractive interaction by the cutoff Cornell potential of \eqnref{Cornellpotn_cutoff}, 
with the cutoff given in \eqnref{rcutoff} and $V_0=0$.

The bound states ($X\bar X$) are characterized by three integers $(l,m,n)$, where $l$ labels the angular momentum, $m$ labels the angular momentum along the $z$-axis, and  $n\geq1$.
An $H_X$--$\bar H_X$ scattering state is approximated by an
\XXb  unbound state, which is characterized by the \XXb relative
momentum $\bfk$, with energy $E_{k}\approx k^2/m_X$, where $k=|\bfk|$.
The reduced mass of both the bound and scattering states is approximately
$m_X/2$.

The bound-state wavefunctions can be written as
\begin{equation}\label{eqn:boundWF}
 \psi_{lm,n}(R_0\bfx)   = \frac {1} {R_0^{3/2}} \frac{\chi_{ln}(x)}{x}Y_{lm}(\Omega_\bfx) \,,
\end{equation}
where the dimensionless coordinate $x$ is defined below \eqnref{radialred}.
The scattering-state wavefunctions can be expanded as
\begin{equation}\label{eqn:unboundWF}
 \phi_{\bfk}(R_0\bfx) = \sum_{l=0}^{\infty}(2l+1)\phi_{\bfk,l}(R_0\bfx)
=\sum_{l=0}^{\infty}(2l+1)\frac{\chi_{kl}(x)}{x}P_l({\hat{\bfk}\cdot\hat{\bfx}})\,.
\end{equation}
For $\bfk$ along $\hat{z}$ this simplifies to
\begin{equation}\label{eqn:unboundWFforZ}
 \phi_{k\hat\bfz,l}(R_0\bfx)=
\sqrt{\frac{4\pi}{2l+1}}
\frac{\chi_{kl}(x)}{x}
Y_l^0(\Omega_\bfx)\,.
\end{equation}
The radial wavefunctions  $\chi_{ln}(x)$ and $\chi_{kl}(x)$ solve
\beq \label{eqn:redSchro}
-\chi''_{ln\,(kl)}(x) + V\w{eff}^l(x) \chi_{ln\,(kl)}(x) = \epsilon_{ln\,(k)}
 \chi_{ln\,(kl)}(x)\,,
\eeq
where
\begin{equation}\label{eqn:V(x)}
\begin{split}
 V\w{eff}^l(x) &= \frac{l(l+1)}{x^2} + V(x) 
 \\&=
 \frac{l(l+1)}{x^2} + \left[{-a_D}\left(\frac{1}{x}-\frac{1}{\xc}\right) + \left(x-\xc\right)  \right] \Theta(\xc-x) \,.
\end{split}
\end{equation}
Note that this dimensionless problem has only two parameters: $x_c$ and the effective Coulomb strength $a_D\equiv(m_X/\Lambda_D)^{2/3} \bar{\alpha}_D$.
The radial wavefunctions are zero at the origin, $\chi_{ln\,(kl)}(0) = 0$, and satisfy the normalization conditions
\begin{equation}\label{eqn:WFNormalization}
\int_0^\infty\chi_{ln}(x)\chi^*_{ln'}(x) dx = \delta_{nn'} \,, \qquad
\int_0^\infty\chi_{kl}(x)\chi^*_{k'l}(x) dx = \frac{\pi}{2k^2}\delta(k-k') \,,
\end{equation}
so that
$\braket{\psi_{lm,n}|\psi_{l'm',n'}}=\delta_{ll'}\delta_{mm'}\delta_{nn'}$
and
$\braket{\phi_{\bfk}|\phi_{\bfk'}}=(2\pi)^3\delta^{(3)}(\bfk-\bfk')$.

\begin{figure}[t]
\centering
\begin{subfigure}[b]{0.49\textwidth}
 \includegraphics[scale=0.72]{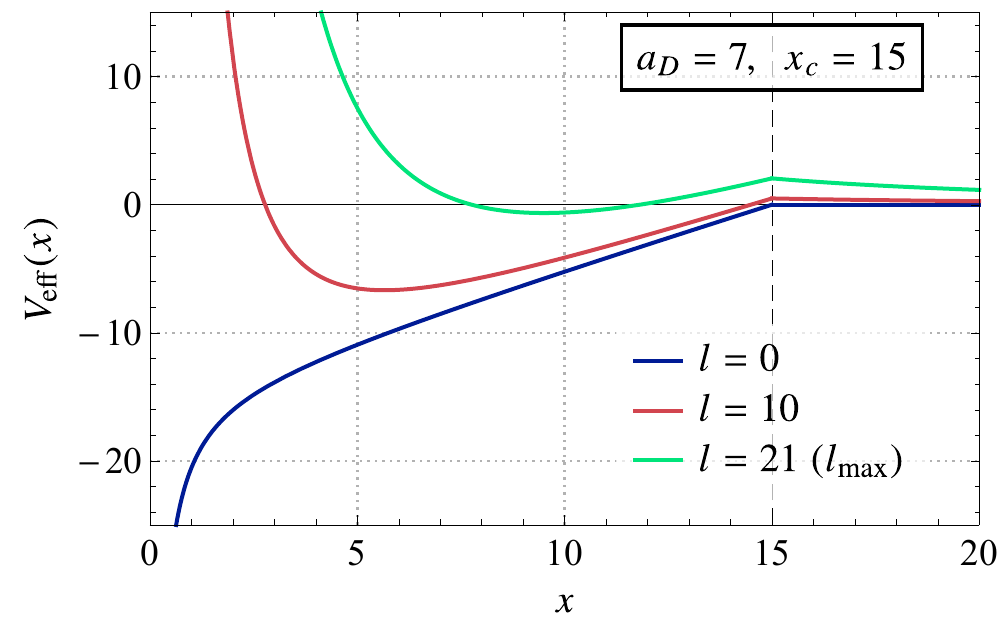}
 \caption{$a_D=7$}
 \label{fig2:VeffCornell}
\end{subfigure}
\hfil
\begin{subfigure}[b]{0.49\textwidth}
 \includegraphics[scale=0.72]{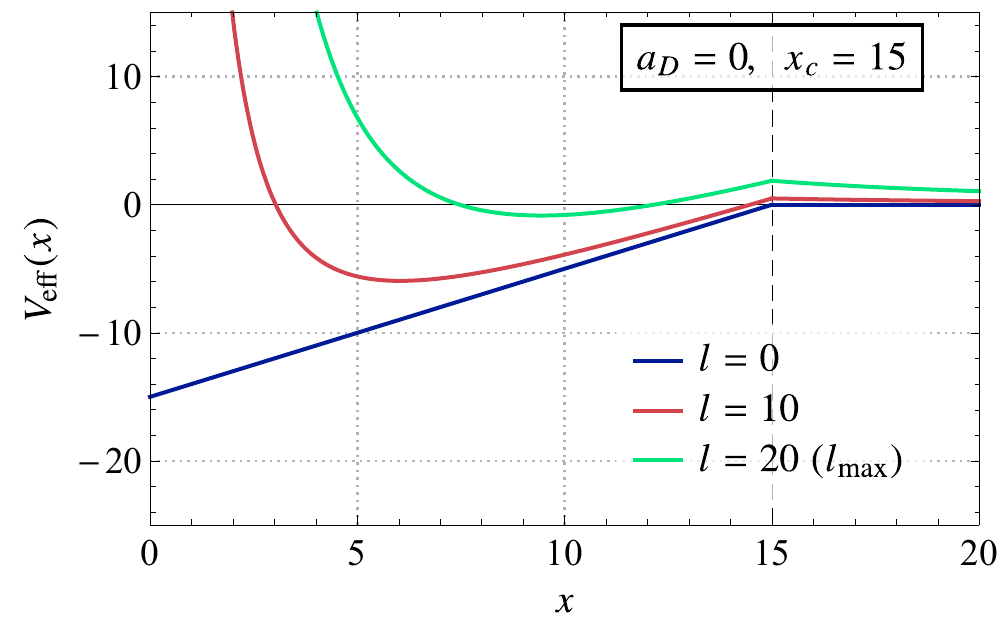}
 \caption{$a_D=0$ (linear)}
 \label{fig2:VeffLinear}
\end{subfigure}
\caption{
The effective potential $V\w{eff}(x)$ with $\xc=15$ $(m_X\simeq125\Lambda_D)$, for three values of $l$.
In the left figure, $a_D=7$, which corresponds to $\bar{\alpha}_D=0.3$, while $a_D=0$ in the right figure.
The largest $l$ in each figure corresponds to the upper-bound on $l$ of the bound states.
Note that the plots for $l=0$ correspond to $V(x)$ in \eqnref{V(x)}.
}
\label{fig2:Veff}
\end{figure}

Figure~\ref{fig2:Veff} shows $V\w{eff}(x)$ in \eqnref{V(x)} for $\xc=15$ with $a_D=7$ (left) and $a_D=0$ (right).
Because of the cutoff, the angular momentum quantum number $l$ of bound states has an upper bound  $l_\max$ given in \eqnref{lmax}.
This is confirmed in the numerical results.\footnote{%
In fact, as one can see in Fig.~\ref{fig2:Veff}, the effective potential for very large $l$ may have a minimum with $V\w{eff}(x_0)\ge 0$,
and it produces wavefunctions  with $\epsilon>0$ which are mostly confined
to the region $x<\xc$.
}

\begin{figure}[t] 
\centering
\begin{subfigure}[b]{0.49\textwidth}
 \includegraphics[scale=0.72]{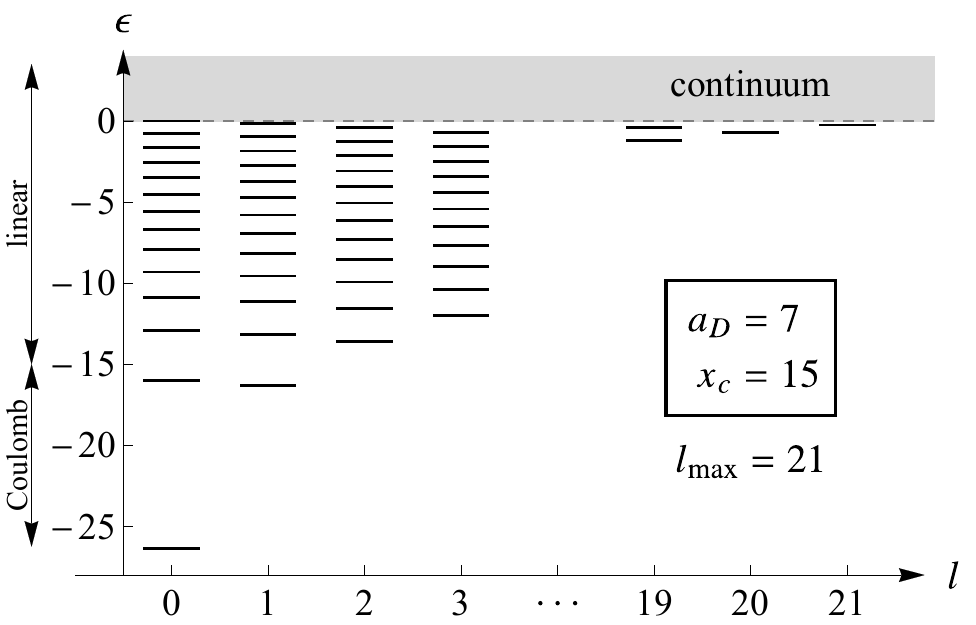}
 \caption{$a_D=7$}
 \label{fig:ElevelsCornell}
\end{subfigure}
\hfil
\begin{subfigure}[b]{0.49\textwidth}
 \includegraphics[scale=0.72]{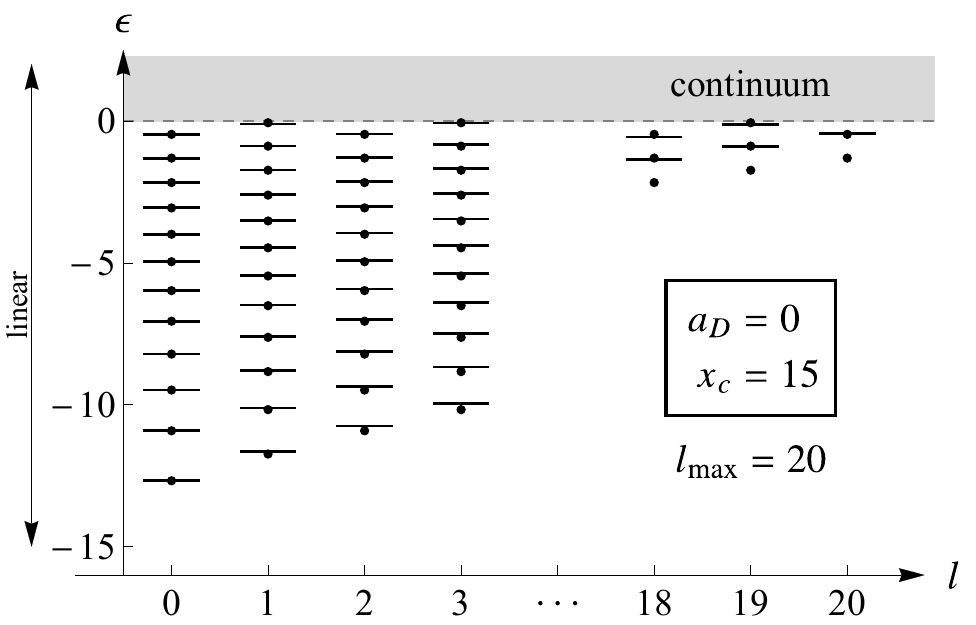}
 \caption{$a_D=0$ (linear)}
 \label{fig:ElevelsLinear}
\end{subfigure}
\caption{
The energy spectrum of bound states. 
We also show, for the linear potential, the semiclassical results in \eqnref{BSenergies} as the small dots.
}
\label{fig:Elevels}
\end{figure}

The bound-state energy levels are shown in Fig.~\ref{fig:Elevels} for a cutoff Cornell (left) and linear (right) potentials.
In Ref.~\cite{Quigg:1979vr}, the semiclassical approximation is used to obtain the energy levels of the eigenstates of central potentials. Following their discussion, the energy levels under the linear potential are given by
\begin{equation}
E_{ln} - V(0) \approx \left[ \frac32 \pi \left(n + \frac{l}2 - \frac14\right) \right]^{2/3}E_0\,,
\end{equation}
or in our notation, since $V(0)=-R_c \Lambda^2_D=-x_c E_0$,
\begin{equation}
 \label{eqn:BSenergies}
 \e_{ln} \approx \left[ \frac32 \pi \left(n + \frac{l}2 - \frac14\right) \right]^{2/3} - x_c\,.
\end{equation}
We reproduce this result in Fig.~\ref{fig:ElevelsLinear}.
For a Cornell potential (Fig.~\ref{fig:ElevelsCornell}), deep states are governed by the Coulomb force 
and therefore obey the well-known Coulombic energy levels
\begin{equation}
 \epsilon_{ln} \approx -\frac{a_D^2}{4 N^2} - \xc\,,
\end{equation}
where the principal quantum number $N$ is given by $n+l$. 
For shallower states, the Coulomb force is negligible and the linear potential governs the spectrum.\footnote{%
The average size $\mean{x}$ of bound states is given by the virial theorem as
$
\epsilon+\xc=-{a_D}/(2\mean{x})
$
for a Coulombic bound states, i.e., if the effect of the linear term is negligible,
and
$
 \epsilon+\xc=3\mean{x}/2
$
for bound states in the linear regime.
}

\begin{figure}[t] 
\centering
\includegraphics[scale=0.72]{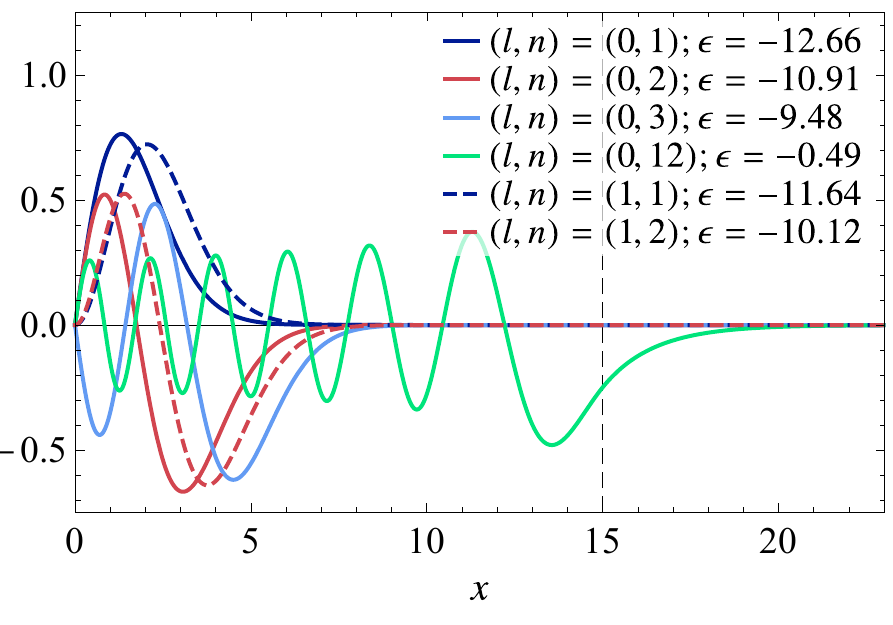}\hfil
\includegraphics[scale=0.72]{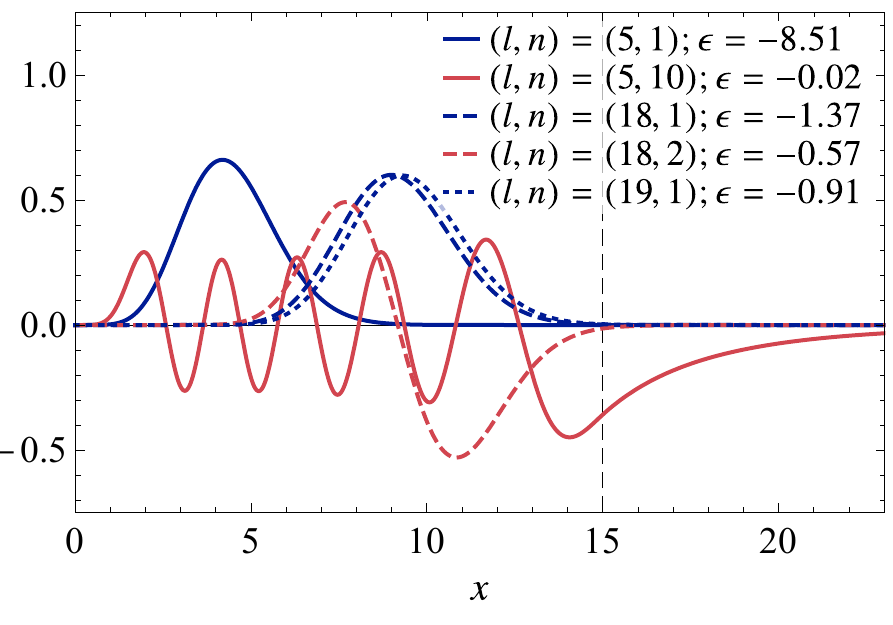}
\vspace{0.5em}
\includegraphics[scale=0.72]{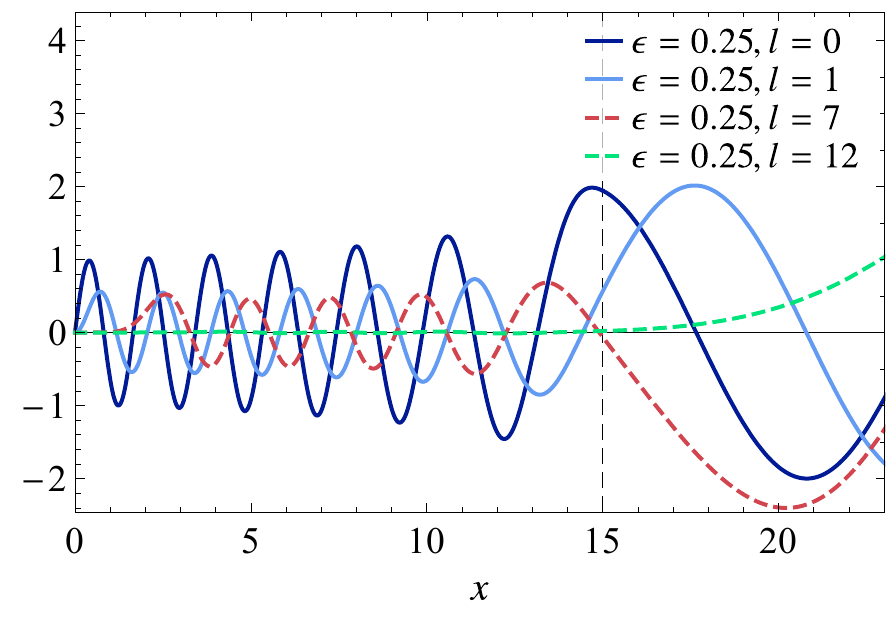}\hfil
\includegraphics[scale=0.72]{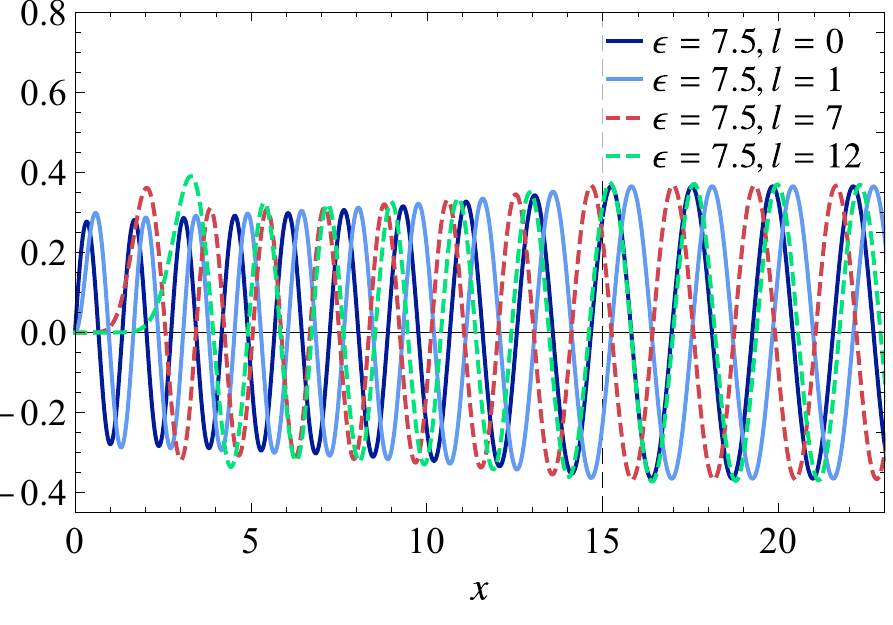}
\caption{
Wavefunctions obtained from solving Eqs.~(\ref{eqn:redSchro}--\ref{eqn:V(x)}) with $\xc=15, a_D=0$. \\
\emph{Top:} $\chi_{ln}(x)$ for bound states.
The left figure contains smaller $l$ (0,1) while the right figure has intermediate and high $l$ (5, 18, $19=l\w{max}$), 
where $(l,n)=(5,10)$ is the highest-energy bound state. \\
\emph{Bottom:} $\chi_{kl}(x)$ for scattering states.
The left figure displays the partial waves of the scattering states with $\epsilon=0.25$ ($E_k=0.05\Lambda_D$),
while the right figure shows those with $\epsilon=7.5$ ($E_k=1.5\Lambda_D$).
}
\label{fig:wfboundscatt}
\end{figure}


We show some examples of bound-state and scattering-state wavefunctions that solve \eqnref{redSchro} with the cutoff linear potential (\eqnref{V(x)} with $a_D=0$) in the top and bottom rows of \figref{wfboundscatt}, respectively.
The bound state with $(l,n)=(5,10)$ is found to be the shallowest bound state, but it should be emphasized that 
this is an accident due to the cutoff being just above the energy of this state. 
In general, shallower states with smaller $l$ have wavefunctions that tend to penetrate beyond $x>\xc$.
For the scattering states, wavefunctions of states with larger $\epsilon$ and smaller $l$ have penetrate further into the region $x<x_c$.


\subsection{Bound-state formation cross section in the dipole approximation} \label{app:bsfxs}

We calculate the matrix element for the bound-state formation by a vector-mediated interaction,
\begin{equation}
 H_X + \bar{H}_X \to (X\bar{X})_{lm,n} + \varphi
\end{equation}
with the $X$--$\varphi$ interaction given by
\begin{equation}
\mathcal L\ni \left\lvert\left(\partial_\mu - i e_X \varphi_\mu\right)X\right\rvert^2 \,.
\end{equation}
We follow the approach and notation in Refs.~\cite{Petraki:2015hla,Petraki:2016cnz}.
From now on, we focus on the linear potential and set $a_D=0$ because, as we will see, 
the radiative bound-state formation process favors shallow bound states, for which the Coulomb force is negligible.

The bound-state formation cross section is given by~\cite{Petraki:2015hla}
\beq
(v\w{rel}\sigma)^{\mathrm{BSF}}_{\bfk}
 = \sum_{l,m,n}
(v\w{rel}\sigma)^{\mathrm{BSF}}_{\bfk\to lm,n}
=\sum_{l,m,n}
\int\frac{d\Omega_{\bfPphi}}{4\pi}
\frac{|\bfPphi|}{32\, \pi\, m^3_X}
\sum\w{pol.}
\left| \bfe\cdot\boldsymbol{\mathcal{M}}_{\bfk\to lm,n} \right|^2
\eeq
in the CM frame, where $\bfk$, $\bfPphi$, and $\bfe$ are the relative momentum of the initial state, the momentum of the radiated light state $\varphi$, and its polarization vector, respectively, and
\begin{equation}
 v\w{rel} \simeq \frac{|\bfk|}{m_X/2}\,,
\qquad
|\bfPphi| = E_k-E_{ln}\,;
\end{equation}
$E_k$ is kinetic energy of the initial scattering state.
The matrix element is
\begin{equation}
 \mathcal M^j_{\bfk\to lm,n}
 =-4e_X\sqrt{m_X}\intdMom3{\bfp}\,p^j\tilde\psi_{lm,n}^*(\bfp)
 \left[
 \tilde\phi_{\bfk}\left(\bfp+\frac{\bfPphi}{2}\right)
 +\tilde\phi_{\bfk}\left(\bfp-\frac{\bfPphi}{2}\right)
 \right],
\end{equation}
where $\tilde{\f}_{\bfk} (\bfp)$ and $\tilde{\j}_{lm,n}(\bfp)$ are the momentum-space equivalents of Eqs.~\eqref{eqn:boundWF}--\eqref{eqn:unboundWF}.

It is convenient to expand the matrix element in partial waves (cf.~\eqnref{unboundWF}):
\begin{align*}
\mathcal M^j_{\bfk\to lm,n} &= \sum_{l'=0}^\infty (2l'+1)\mathcal M^j_{\bfk,l'\to lm,n} \,, \\
 \mathcal M^j_{\bfk,l'\to lm,n} &=
-4e_X\sqrt{m_X}
\intdMom3{\bfp}\,p^j\tilde\psi_{lm,n}^*(\bfp)
 \left[
 \tilde\phi_{\bfk,l'}\left(\bfp+\frac{\bfPphi}{2}\right)
 +\tilde\phi_{\bfk,l'}\left(\bfp-\frac{\bfPphi}{2}\right)
 \right] \notag\\
&=
8i e_X\sqrt{m_X}\int\!\!d^3\bfr\,d^3\bfr'\,
\psi^*_{lm,n}(\bfr)
\f_{\bfk,l'}(\bfr^\prime) \cos\left( \frac{\bfPphi}2 \cdot \bfr^\prime \right)
\frac{\partial}{\partial r^j} \delta^{(3)}(\bfr - \bfr^\prime) \,. \notag
\end{align*}
After partial integration, the derivative acts on $\j^*_{lm,n}(\bfr)$,
and the integral over $d^3\bfr^\prime$ is trivial using the delta function.
Dotting with the polarization vector and performing another partial integration (the derivative of the cosine is zero as $\bm{\e} \cdot \bfPphi = 0$) yields
\beq \label{eqn:polmatelem}
\bfe\cdot \bfcalM_{\bfk,l'\to lm,n}
= 4ie_X \sqrt{m_X} \int d^3\bfr \, \cos\left( \frac{\bfPphi}2 \cdot \bfr \right)
\bm{\e} \cdot \left[ \j^*_{lm,n} \bm{\nabla} \f_{\bfk,l'} - \f_{\bfk,l'} \bm{\nabla} \j^*_{lm,n} \right] \,,
\eeq
where $\j^*_{lm,n}$ and $\f_{\bfk,l'}$ are functions of $\bfr$.
The quantity in brackets above appears in the difference of the Schr\"odinger equations for the scattering- and bound-state wavefunctions,
\beq
-\frac1{m_X} \bm{\nabla} \cdot (\j^*_{lm,n} \bm{\nabla} \f_{\bfk,l'} - \f_{\bfk,l'} \bm{\nabla} \j^*_{lm,n}) = (E_k - E_{nl}) \j^*_{lm,n} \f_{\bfk,l'} \,.
\eeq
Using the identity%
\footnote{By assumption, $\bfF$ satisfies appropriate fall-off behavior at large $|\bfr|$ so the boundary term is negligible.}
\beq
\int d^3\bfr \, \left( {\nabla} \cdot \bfF\right) \bfr = -\int d^3\bfr \, \bfF \left( \bm{\nabla} \cdot \bfr\right) = -\int d^3\bfr \, \bfF
\eeq
with $\bfF = \j^*_{lm,n} \bm{\nabla} \f_{\bfk,l'} - \f_{\bfk,l'} \bm{\nabla} \j^*_{lm,n}$
and substituting into \eqnref{polmatelem}, we obtain%
\footnote{This is the form of the matrix element in Refs.~\cite{Wise:2014jva, An:2016gad}; both employ the dipole approximation and the Hamiltonian formulation.}
\beq
\begin{split}
\bfe\cdot \bfcalM_{\bfk,l'\to lm,n}
= 4ie_X \sqrt{m_X} & \int d^3\bfr \, (\bfe \cdot \bfr)  \Bigg[
m_X (E_k - E_{ln}) \j^*_{lm,n} \f_{\bfk,l'} \cos\left( \frac{\bfPphi}2 \cdot \bfr \right) \\
& + \frac{\bfPphi}2 \cdot (\j^*_{lm,n} \bm{\nabla} \f_{\bfk,l'} - \f_{\bfk,l'} \bm{\nabla} \j^*_{lm,n}) \sin\left( \frac{\bfPphi}2 \cdot \bfr \right) \Bigg] \,.
\end{split}
\eeq

We are interested in temperatures $T\lesssim\Lambda_D$ for which $\bfPphi\ll m_X$, and thus
\beq\label{eqn:beforedipole}
\bfe\cdot \bfcalM_{\bfk,l'\to lm,n}
\simeq 4ie_X \sqrt{m_X^3} (E_k - E_{ln}) \int d^3\bfr \,
(\bfe \cdot \bfr)
\j^*_{lm,n} \f_{\bfk,l'} \cos\left( \frac{\bfPphi}2 \cdot \bfr \right).
\eeq
Also, we can evaluate $\bfPphi\cdot\bfr/2$ as
\begin{equation}
\frac{\bfPphi}{2}\cdot\bfr
\le
\frac{|\bfPphi|r}{2}
=
\frac{|E_k-E_{ln}|r}{2}
=
\kappa_{\Lambda_D} \frac{|E_k-E_{ln}|}{\Lambda_D} \frac{r}{R_c}\,.
\end{equation}
Note that as the integrand contains a bound-state wavefunction, the integral has support only for $r\lesssim R_c$.
Also, for $T\ll\Lambda_D$, the overlaps of the bound and scattering states are larger for shallower bound states.
Combining these, we can approximate $\bfPphi\cdot\bfr/2\ll1$
in order to employ the dipole approximation $\cos(\bfPphi\cdot\bfr/2)\to 1$, which
simplifies the cross section to
\begin{align}
 (v\w{rel}\sigma)^{\mathrm{BSF}}_{\bfk}
 &\to
 \sum_{l,m,n}
 \int\frac{d\Omega_{\bfPphi}}{4\pi}
 \frac{|\bfPphi|}{32\, \pi\, m^3_X}
 \sum\w{pol.}
 \left|
 \sum_{l'}(2l'+1)
 4ie_X \sqrt{m_X^3} (E_k - E_{ln})
 \bfe\cdot{\boldsymbol{\mathcal{I}}}_{\bfk,l^\prime \to lm,n}
 \right|^2
 \notag\\&=
 \sum_{l,m,n}
 \int\frac{d\Omega_{\bfPphi}}{4\pi}
 \frac{e_X^2(E_k - E_{ln})^3}{2\pi}
 \sum\w{pol.}
 \left|
 \sum_{l'}(2l'+1)
 \bfe\cdot{\boldsymbol{\mathcal{I}}}_{\bfk,l^\prime \to lm,n}
 \right|^2,
 \label{eqn:radXS}
\end{align}
where the integral of the wavefunctions is defined as
\beq \label{eqn:Ivec}
\begin{split}
\mathcal{I}^j_{\bfk,l^\prime \to lm,n}
=  \int d^3\bfr \, r^j \, \j^*_{lm,n} \, \f_{\bfk,l'} \,.
\end{split}
\eeq

Next, we express the integral in the basis $r^a=(r^+,r^0,r^-)$ defined by
\begin{equation}\label{eqn:polBasis}
r^{\pm} = \frac{-1}{\sqrt{2}} \left(\pm r^1 + i r^2\right) = r \, \sqrt{\frac{4\pi}3} \, Y^{\pm1}_1\left(\O_\bfr\right) \,, \qquad
r^0 = r \, \sqrt{\frac{4\pi}3} \, Y^0_1\left(\Omega_\bfr\right) \,.
\end{equation}
Taking the $z$-axis parallel to $\hat\bfk$, substituting \eqnref{boundWF} and \eqnref{unboundWFforZ} for the wavefunctions in \eqnref{Ivec},
we obtain
\begin{align}
\mathcal{I}^a_{\bfk,l^\prime \to lm,n}
&=
\frac{4\pi R_0^{5/2}}{\sqrt{3(2l'+1)}}
I_{kl'\to ln}
\int d\Omega_\bfx Y_{l'}^0(\Omega_\bfx)Y^{a}_1(\Omega_\bfx)Y_l^{m*}(\Omega_\bfx)
\notag\\&=
(-1)^m
\sqrt{{4\pi(2l+1)}}
\,
R_0^{5/2}I_{kl' \to ln}
\ThreeJzero{l'}{1}{l}
\ThreeJ{l'}{1}{l}{0}{a}{-m}
,\label{eqn:IvecIn3Jform}\\
I_{kl' \to ln}
&= \int dx \, x \, \chi^*_{ln}(x)\,\chi_{kl^\prime}(x) \,,
\end{align}
where we have expressed the integral over solid angle of three spherical harmonics in terms of Wigner $3j$-symbols
\beq\label{eqn:3Yintegral}
\int d\O \, Y_{l_1}^{m_1} Y_{l_2}^{m_2} Y_{l_3}^{m_3} =
\left[ \frac{(2l_1+1) (2l_2+1) (2l_3+1)}{4\pi} \right]^{1/2}
\ThreeJzero{l_1}{l_2}{l_3}
\ThreeJ{l_1}{l_2}{l_3}{m_1}{m_2}{m_3} \,.
\eeq
These symbols give the selection rules
$l' = |l \pm 1|$ and $m = a$,
so that $|m|\le 1$.

With these results, the sum over $m, l^\prime$, and polarizations in \eqnref{radXS} evaluates to%
\footnote{Note that $\mathcal I^a_{k\hat\bfz,l-1\to lm,n}$ is always zero if $l=0$.}
\begin{align}
&\sum_{m=-l}^l
\sum\w{pol.}
\left|
\sum_{l'}(2l'+1)
\bfe\cdot{\boldsymbol{\mathcal{I}}}_{\bfk,l^\prime \to lm,n}
\right|^2
\notag\\&=
\sum_{m=-l}^{l}
\sum_{a=-1}^{1}
\left|
 (2l+3) {\mathcal{I}}^m_{\bfk,l+1 \to lm,n}
+
 (2l-1) {\mathcal{I}}^m_{\bfk,l-1 \to lm,n}
\right|^2\delta_{am}
\notag\\&=
\sum_{m=-1}^{1}
\left|
 (2l+3) {\mathcal{I}}^m_{\bfk,l+1 \to lm,n}
+
 (2l-1) {\mathcal{I}}^m_{\bfk,l-1 \to lm,n}
\right|^2
\notag\\&=
4\pi R_0^5
\left[
(l+1)
\left|
I_{k,l+1 \to ln}
\right|^2
+
l\left|
I_{k,l-1 \to ln}
\right|^2
\right].
\end{align}
Inserting this into the cross section in \eqnref{radXS} and trivially performing the $d\Omega_{\bfPphi}$ yields
\begin{align}
 (v\w{rel}\sigma)^{\mathrm{BSF}}_{\bfk}
&= 2e_X^2R_0^5\sum_{l,n}(E_k - E_{ln})^3
\left[
(l+1)\left|I_{k,l+1 \to ln}\right|^2
+ l\left|I_{k,l-1 \to ln}\right|^2
\right]
\notag\\&=
 \frac{2e_X^2}{m_X^2}\left(\frac{\Lambda_D}{m_X}\right)^{2/3}
 \sum_{l,n}J_{k,ln} \,,
\label{eqn:dipoleradxsresult}
\end{align}
with $J_{k,ln}$ defined in \eqnref{J}.


\subsection{Thermally-averaged cross section} \label{app:thermalxs}

Here we briefly describe the procedure to calculate the thermally-averaged cross section, 
$\mean{v\w{rel}\sigma}(\beta)$, as a function of the inverse temperature $\beta=T^{-1}$.
As in the previous discussion, we denote the momenta of the initial particles as $\bfk_1$, $\bfk_2$.

As we are interested in $T\ll m_X$, the kinetic distributions of $H_X$ and $\bar{H}_X$ are given by the Maxwell-Boltzmann distribution,
\begin{equation}
 f\w{MB}(\bfp) = \fracp{2\pi\beta}{m}{3/2}\exp\left(-\frac{\beta |\bfp|^2}{2m_X}\right),
\end{equation}
which is normalized so that $\intdMom3{\bfp}f\w{MB}(\bfp)=1$.
With this distribution, the thermally-averaged cross section is given by
\begin{align}
\mean{v\w{rel}\sigma}(\beta)
&=
\intdMom3{\bfk_1}
\dMom3{\bfk_2}
f\w{MB}(\bfk_1) f\w{MB}(\bfk_2)
(v\w{rel}\sigma)^{\mathrm{BSF}}_{\bfk}
\notag\\&=
\fracp{2\pi\beta}{m_X}{3}
\intdMom3{\bfK}
\dMom3{\bfk}
\exp\left(-\frac{\beta(|\bfK/2|^2+|\bfk|^2)}{m_X}\right)
(v\w{rel}\sigma)^{\mathrm{BSF}}_{\bfk}
\notag\\&=
\sqrt{\frac{16\beta^3}{\pi m_X^3}}
\int\!\!dk\,
 k^2\,e^{-\beta k^2/m_X}
(v\w{rel}\sigma)^{\mathrm{BSF}}_{k}\,,
\end{align}
where, as before, $\bfK=\bfk_1+\bfk_2$ and $\bfk=(\bfk_1-\bfk_2)/2$.
We have neglected the dependence of the (non-relativistic) cross section 
$(v\w{rel}\sigma)^{\mathrm{BSF}}_{k}$ on $\bfK$ since $|\bfK|\ll|\bfk|$.

Combining the previous result and also including the final-state effect of $\varphi$, we obtain
\begin{align}
 \mean{v\w{rel}\sigma}(\beta)
=
 \frac{2e_X^2}{m_X^2}\fracp{\Lambda_D}{m_X}{2/3}
\int\!\!dk\,w(k;\beta)
 \sum_{l,n}\frac{J_{k,ln}}{1-\exp\left[-\beta(E_k-E_{ln})\right]}\,,
\end{align}
where we define
\begin{equation}
w(k;\beta) = \sqrt{\frac{16\beta^3}{\pi m_X^3}}\,k^2\,e^{-\beta k^2/m_X}\,, \qquad \int w(k;\beta)dk=1 \,.
\end{equation}


\bibliography{bibliography_su(n)-dm_jhep}

\end{document}